\documentclass[]{rsos}%%%%where rsos is the template name

\usepackage{amssymb}
\usepackage{amsmath}
\usepackage{color}
\usepackage{bm}
\usepackage{hyperref}
\usepackage{graphicx}
\usepackage{harvard}
\usepackage{epsfig}
\usepackage{longtable}
\usepackage{lscape}
\usepackage{url}
\usepackage{threeparttable}
\usepackage[svgnames]{xcolor}

\usepackage{ulem}

\def\new#1 {{\bf #1 }}

\def\kmspc {km\,s$^{-1}$\,pc$^{-1}$}
\def\kms {km\,s$^{-1}$}
\def\Htwo {H$_2$}

\def\Kkms {{\sc k}\,km\,s$^{-1}$}
\def\dvdr {${\rm d}v/{\rm d}r$}

\def\Td {$T_{\rm d}$ }

\def\SFRsd {$M_{\rm \odot}$\,yr$^{-1}$\,kpc$^{-2}$}

\def\purple#1 {{\textcolor{purple}{#1}}\ }
\def\red#1 {\textcolor{red}{#1}}
\def\new#1 {{\bf #1 }}

\begin{document}

%%%% Article title to be placed here
\title[CMB effect in imaging distant galaxies]{Gone with the heat: A fundamental constraint on the imaging of dust and molecular gas in the early Universe}

\date{\today}

\author{\Large Zhi-Yu\,Zhang$^{1,2}$, Padelis\,P.\,Papadopoulos$^{3,2,4}$, R.\,J.~Ivison$^{2,1}$, Maud Galametz$^{2}$, M.\,W.\,L.\ Smith$^{3}$ and Emmanuel M.\ Xilouris$^{5}$}

\address{\small{\noindent$^{1}$Institute for Astronomy, University of Edinburgh, Royal Observatory, Blackford Hill, Edinburgh EH9 3HJ, UK\\
$^{2}$ESO, Karl-Schwarzschild-Str.~2, D-85748 Garching,  Germany\\
$^{3}$School of Physics and Astronomy, Cardiff University, Queen's Buildings, The Parade, Cardiff CF24 3AA, UK \\ 
$^{4}$Research Center for Astronomy, Academy of Athens, Soranou Efesiou 4, GR-115 27 Athens, Greece \\
$^{5}$Institute for Astronomy, National Observatory of Athens, GR-15236 Penteli, Greece}}

%%%% Subject entries to be placed here %%%%
\subject{Astrophysics, Cosmology, Interstellar medium} 

%%%% Keyword entries to be placed here %%%%
\keywords{Galaxies: ISM, Cosmology: cosmic background radiation, Galaxies: high-redshift}

%%%% Insert corresponding author and its email address}
\corres{Zhi-Yu Zhang\\
\email{zzhang@eso.org}}

%%%% Abstract text to be placed here %%%%%%%%%%%%
\begin{abstract}
Images of dust continuum and carbon monoxide (CO) line emission are powerful tools for deducing
structural characteristics of galaxies, such as disk sizes, \Htwo\ gas velocity
fields and enclosed \Htwo\ and dynamical masses. We report on a fundamental
constraint set by the cosmic microwave background (CMB) on the observed
structural and dynamical characteristics of galaxies, as deduced from dust
continuum and CO-line imaging at high redshifts. As the CMB temperature rises
in the distant Universe, the ensuing thermal equilibrium between the CMB and
the cold dust and \Htwo\ gas progressively erases all spatial and spectral
contrasts between their brightness distributions and the CMB. For high-redshift
galaxies, this strongly biases the recoverable \Htwo\ gas and dust mass
distributions, scale lengths, gas velocity fields and dynamical mass estimates.
This limitation is unique to mm/submm wavelengths and unlike its known effect
on the global dust continuum and molecular line emission of galaxies, it cannot
be addressed simply. We nevertheless identify a unique signature of
CMB-affected continuum brightness distributions, namely an increasing rather
than diminishing contrast between such brightness distributions and the CMB
when the cold dust in distant galaxies is imaged at frequencies beyond the
Raleigh-Jeans limit. For the molecular gas tracers, the same effect makes the
atomic carbon lines maintain a larger contrast than the CO lines against the
CMB.
\end{abstract}
%%%%%%%%%%%%%%%%%%%%%%%%%%%

%%%%%%%%%% Insert the texts which can accomdate on firstpage in the tag "fmtext" %%%%%
\maketitle

\section{Introduction}

Imaging the thermal continuum from cosmic dust and the CO line emission from
its concomitant molecular (H$_2$) gas at mm and submm wavelengths has become a
powerful tool for deducing structural characteristics of galaxies
\cite{Young1995ApJS,Thomas2004MNRAS,Wong2002ApJ,Downes1998ApJ,Daddi2010ApJ,Hodge2012ApJ,Riechers2013}.
Moreover the H$_2$ gas velocity field together with its relative mass
distribution in galactic disks are indispensable for studying instabilities in
the evolution of such systems from the early Universe to the present day
\cite{Daddi2010ApJ,Hodge2012ApJ,Bournaud2015AA,Bournaud2015B}.

As the CMB temperature rises in the distant Universe, the ensuing thermodynamic
interaction between the CMB and the cold interstellar medium (ISM; i.e.\ dust
and H$_2$ gas) becomes important and must be taken into account
\cite{Combes1999AA,Papadopoulos2000ApJ,DaCunha2013ApJ}.  This effect is
prominent at mm and submm wavelengths, close to the blackbody peak of the CMB.
The latter now becomes the dominant thermal pool with which the cold dust and
\Htwo\ gas of distant galaxies interacts, while it also provides the
irreducible background against which dust continuum and molecular line emission
of gas-rich disks must be measured.

The effect of the CMB on the global (i.e.\ spatially integrated) spectral
energy distributions (SEDs) of the dust emission and on the spectral line
energy distributions (SLEDs) of CO line emission in high-redshift galaxies has
been widely studied \cite{Papadopoulos2000ApJ,DaCunha2013ApJ,Combes1999AA}. Its
impact on measuring the total IR luminosity ($\propto $ star-formation rate,
SFR) for dust-enshrouded objects in the distant Universe and the total mass of
dust and H$_2$ is also well characterised \cite{DaCunha2013ApJ}. However an
elevated CMB will also progressively erase the contrasts between the brightness
distributions of dust continuum and molecular line emission emanating from the
cold interstellar medium of distant galaxies. This effect, and its impact on
the recovered morphologies and dynamics of gas-rich disk in the distant
universe have yet to be studied.

Large masses of cold dust and H$_2$ gas $(T_{\rm dust, kin}\sim 15-25\,${\sc k})
are common in local galaxies
\cite{Thomas2004MNRAS,Dunne2001MNRAS,Smith2012ApJ,Papadopoulos1998ApJ}
including the Milky Way \cite{Fixsen1999ApJ}. In spirals their distributions
contain the bulk of the H$_2$ gas and dust, define the total size of their
H$_2$ gas and dust disks and encompass all major star-forming activity
\cite{Wong2002ApJ,Papadopoulos1998ApJ}. The brightness distribution of cold
dust often extends well past the CO-marked H$_2$ gas disk
\cite{Thomas2004MNRAS,Papadopoulos1999ApJ}, while the low-$J$ CO line emission
from cold H$_2$ gas remains the best -- i.e.\ the most mass-inclusive -- probe
of galactic dynamics in metal-rich disks after the H{\sc i} 21-cm line. There
is no evidence that this picture changes much for distant gas-/dust-rich
star-forming galaxies, where extended cold H$_2$ gas reservoirs are now well
established \cite{Ivison2011MNRAS}. This is expected since, despite the fact
that a large fraction of H$_2$ gas and dust mass is involved in star formation
in high-redshift galaxies (and will belong to a warm ISM phase), star formation
remains a globally inefficient process in disks \cite{Daddi2010ApJ,SHI2014},
leaving massive distributions of H$_2$ gas and dust in a cold low-density state
\cite{Tan2014,Bethermin2015}.

Determining scale lengths of gas disks \cite{Ikarashi2015ApJ}, velocity fields
of molecular gas \cite{Glazebrook2013PASA}, enclosed dynamical gas masses
($M_{\rm dyn}$) \cite{Genzel2014,Tan2014}, Tully-Fischer relations, and the
cosmic evolution of gas-rich disk instabilities
\cite{DeBreuck2014AA,Genzel2014} in high-redshift galaxies depends critically
upon recovering the dust and H$_2$ gas distributions, {\it irrespective of
their thermal state}, i.e.\ warm star-forming (SF) vs cold non-SF gas and dust.
The constraint set by the elevated CMB on the recoverable distributions of cold
H$_2$ gas and dust at high redshifts then leads directly to serious biases for
some of the most important characteristics that are to be determined in
exquisite details with Atacama Large Millimeter Array (ALMA) and the Karl G.\
Jansky Very Large Array (JVLA) \cite{alma2015ApJ,Casey15}. The wavelength range
covered by these facilities is where the elevated CMB has its strongest impact,
namely cm, mm and submm wavelengths, placing constraints on the morphological,
structural, and gas velocity field information that can be recovered by
interferometer arrays.

The structure of this paper is as follows. In \S\ref{math}, we present the
fundamental physics of the CMB effects on the dust continuum distributions. In
\S\ref{cmbdust}, we simulate these CMB effects for the dust emission of three
nearby galaxies over different redshifts, and describe the impact on their
recoverable dust continuum brightness distribution. In \S\ref{COmodel}, we
study the effects of the CMB on the emission of molecular gas tracers, namely
CO and C\,{\sc i} transitions, and describe how the CMB effects change their
observed brightness distributions and recoverable gas velocity fields for
galaxies at high redshift. Finally, in \S\ref{summary}, we give a summary and
some concluding remarks.

\section{Effects of the CMB on the observed dust continuum brightness
distributions}\label{math}

We consider the effects of the CMB on the dust continuum emission at different
redshifts. Following \cite{DaCunha2013ApJ}, we keep the same interstellar
radiation field (ISRF), the same intrinsic dust properties (i.e.\ column
density, dust emissivity spectral index $\beta$). The only variable is the CMB
temperature which increases with redshift as $T_{\rm CMB}(z) = T_{\rm CMB}(0)
\times (1+z)$. We assume that the dust optical depth $\tau \ll 1$ at
radio/(sub)mm wavelengths. When the dust, the ISRF and the CMB reach a thermal
equilibrium, the dust emission can be described with a modified blackbody
(MBB):

\begin{equation}
 \int^{\infty} _0 \nu^{\beta} B_{\nu} (T_{\rm d}) d\nu \propto T_{\rm d}^{4+\beta} 
\end{equation}

The modelling includes two major effects --- the CMB dust heating and the CMB
background continuum subtraction. At a given redshift, dust is heated by the
CMB photons to a temperature $T_{\rm d} (z)$:

\begin{equation}
 T_{\rm d} (z) =
 T_{\rm d}(0) \{ 1 + [(1+z)^{4+\beta} -1]
 [\frac{T_{\rm CMB}(0)} {T_{\rm d}(0)}]^{4+\beta}
 \} ^ { 1 / ( 4+\beta) }
\end{equation}

In observed data the background emission is always removed, by one or more
kinds of background subtraction in single-dish observations or by Fourier
spatial filtering in interferometric observations. The radiation temperature at
the rest frequency, $\nu = \nu_{\rm obs} (1+z)$,  in {\it the source local rest
frame} is $J[T_{\rm d}(z), \nu] - J[T_{\rm CMB}(z), \nu]$, where $J(T,
\nu)=(h\nu/k_{\rm B})/[\exp(h\nu/k_{\rm B}T)-1]$ is the Planck radiation
temperature at a frequency $\nu$ and for a temperature, $T$.

For a given galaxy at redshift, $z$, the brightness in the rest frame $B_{\rm
\nu} \propto ({ J[T_{\rm d}(z), \nu] - J[T_{\rm CMB}(z), \nu]})$, so the
brightness ratio (at the same emitting frequency) between a galaxy at redshift
$z$ and the same galaxy at redshift 0 is:

\begin{equation} \label{Rb}
 R_{\rm B} = \frac { J[T_{\rm d}(z), \nu] - J[T_{\rm CMB}(z), \nu] }
 { J[T_{\rm d}^{z=0} , \nu] - J[T_{\rm CMB}^{z=0} , \nu] }
\end{equation}

\begin{figure*}
 \begin{center}
\includegraphics[scale=.80, angle=0]{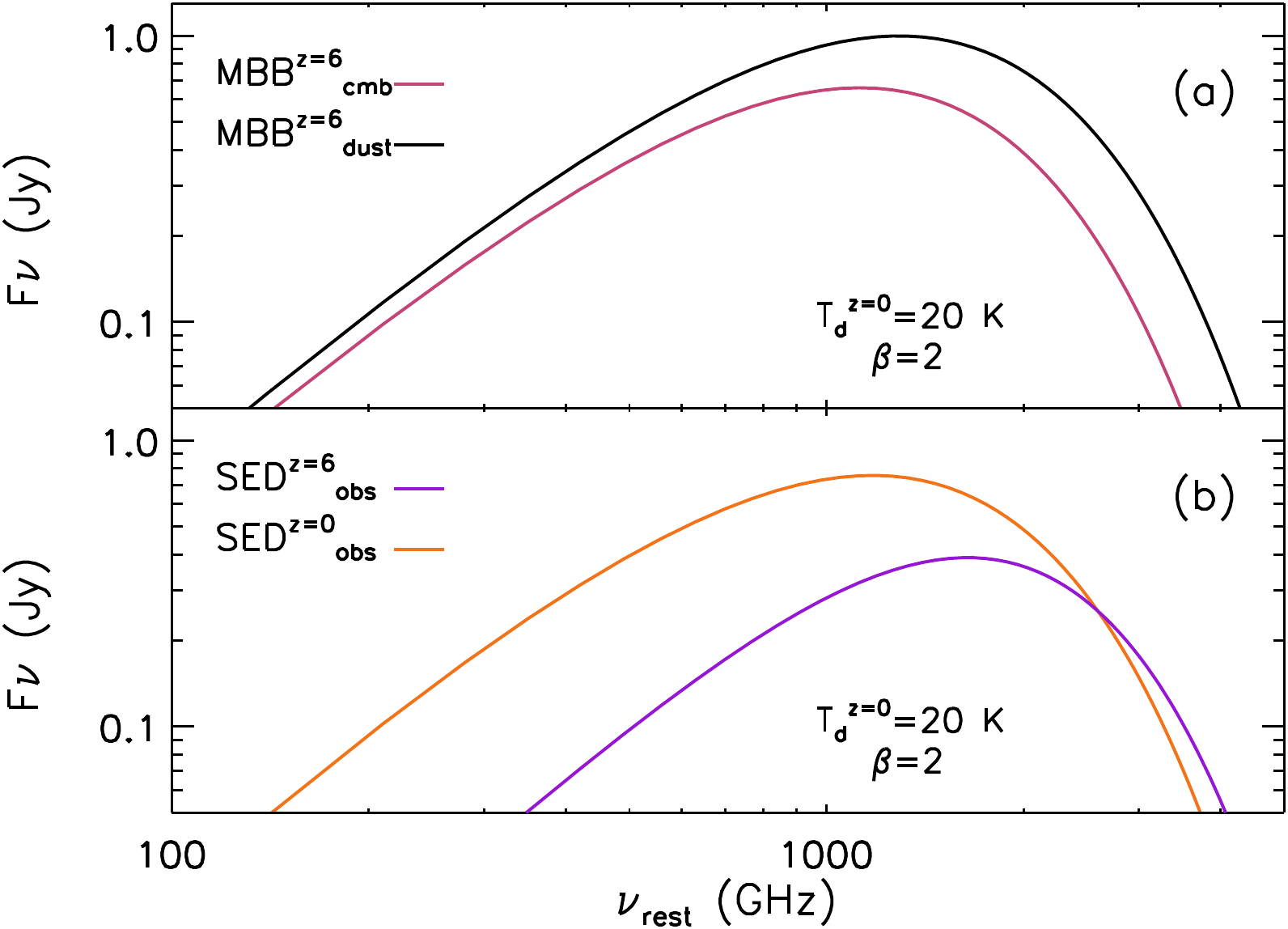}
\caption{Spectral energy distributions (SEDs) of dust with \Td$^{z=0}$  = 20\,{\sc k}, $\beta$ = 2 and at $z
=6$. The upper panel shows the predicted modified black bodies (MBBs) of the CMB emission (red) and the
intrinsic dust emission (black). The lower panel shows the dust SEDs at $z =0$
(brown) and $z =6$ (purple), both after the background subtraction. All plots
are normalised to the peak of the intrinsic dust emission at $z =6$.
}\label{blackbodies}
\end{center}
\end{figure*}

In Fig.~\ref{blackbodies} we investigate the effects of the CMB on dust
emission by comparing the SED at $z = 0$ and $z = 6$. We adopt a local galaxy
with a `normal' cold dust temperature of \Td$^{z=0}$ = 20\,{\sc k} and $\beta =
2$, shift it to redshift $z = 6$, and keep all parameters the same except for
the CMB temperature. The upper panel shows the SEDs of the CMB (red) and the
intrinsic dust MBB (black) at redshift, $z = 6$, and the lower panel shows the
observed rest-frame dust SEDs (after background subtraction) at $z = 0$
(brown) and $z = 6$ (purple).  At low frequencies (the Rayleigh-Jeans domain),
as shown in the lower panel, the dust emission brightness at $z=6$ is lower
than that at redshift 0, with the brightness difference $\delta I_{\nu}/I_{\nu}
\propto \delta T /T $, so that the near-identical temperatures of the dust and
CMB blackbodies would (linearly) translate to near-identical brightness. The
resulting brightness dimming effect will make it difficult to image cold dust
emission distributions at high redshifts at low frequencies.

However we find that beyond a certain frequency the observed contrast against
the source-frame CMB increases again. This is because as the CMB heats up the
dust to higher temperatures and brings their two SEDs ever closer, the
rest-frame frequency crosses over to the Wien side of the two nearly identical
SEDs. Then the resulting brightness difference will be non-linearly boosted
with respect to the underlying small temperature difference between the two
SEDs ($\rm \delta I_{\nu}/I_{\nu} > \delta T/T$), over-compensating for the
dimming due to the ever-closer blackbody functions of the cold dust and the CMB
(see the upper panel of Fig.~\ref{blackbodies}).

An equivalent view of this effect using a galaxy at a given high redshift
(rather than a galaxy `moving' out to progressively higher redshifts) can be
recovered if we consider ever higher frequency observations of its cold dust
continuum brightness distribution. Then, as long as the imaging observations
are performed at a frequency high enough to have the source-frame frequency
cross from the Rayleigh-Jeans to the Wien side of the cold dust distribution,
the observed dust continuum distribution will re-brighten. {\it This effect can
then serve as a clear indicator, showing that a low-brightness dust continuum
distribution observed at low frequencies in a distant galaxy is due to the CMB
bias rather than to low dust mass surface densities.} This re-brightening
effect will typically occur when we observe a high-redshift cold dust disk at
relatively high observing frequencies --- from the high end of the submm to the
THz regime, and it could even permit the recovery of its cold dust mass
distribution, which may be impossible for frequencies in the Rayleigh-Jeans
domain.

\begin{figure*}
 \begin{center}
\includegraphics[scale=.35]{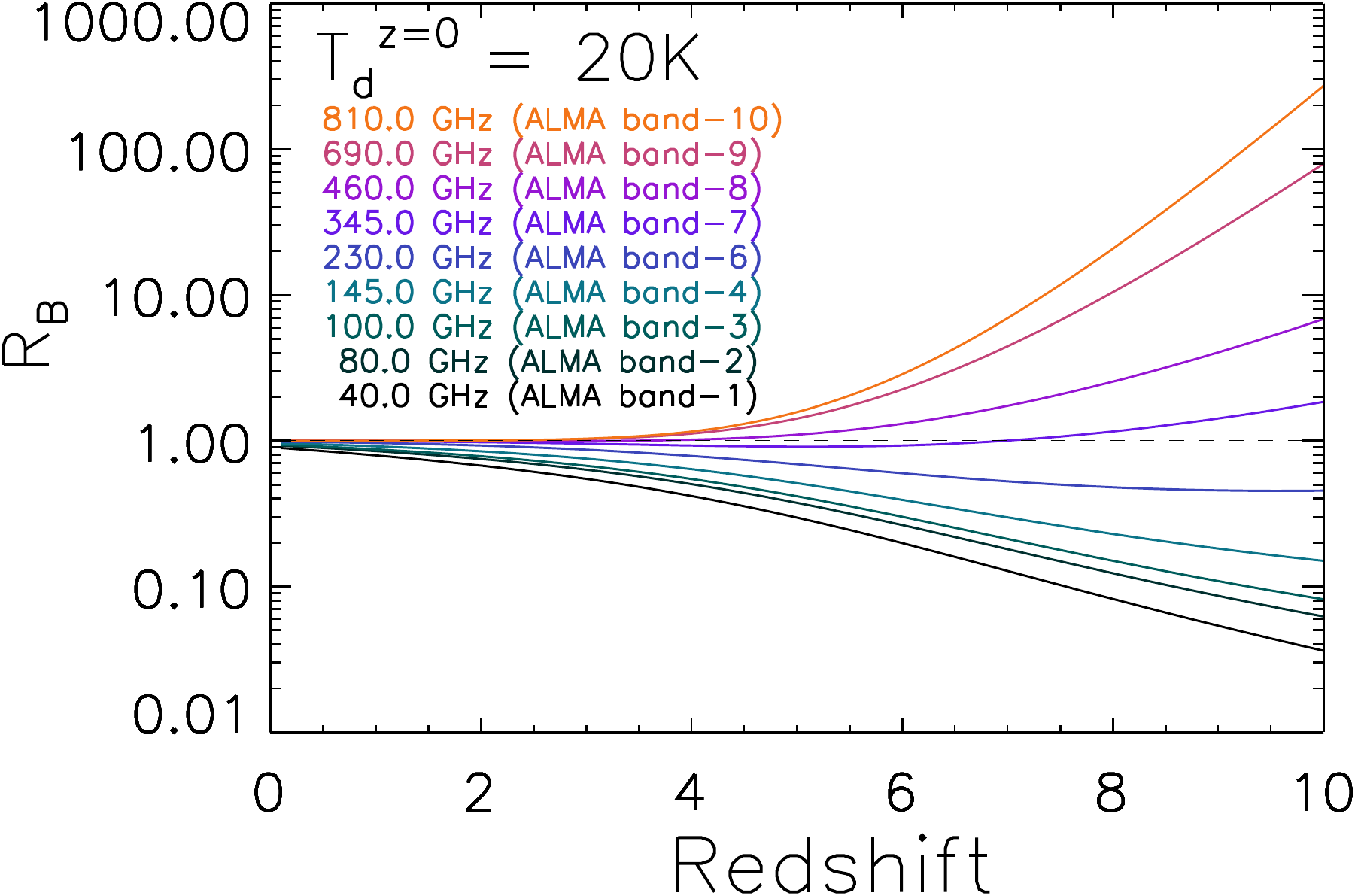} 
\includegraphics[scale=.35]{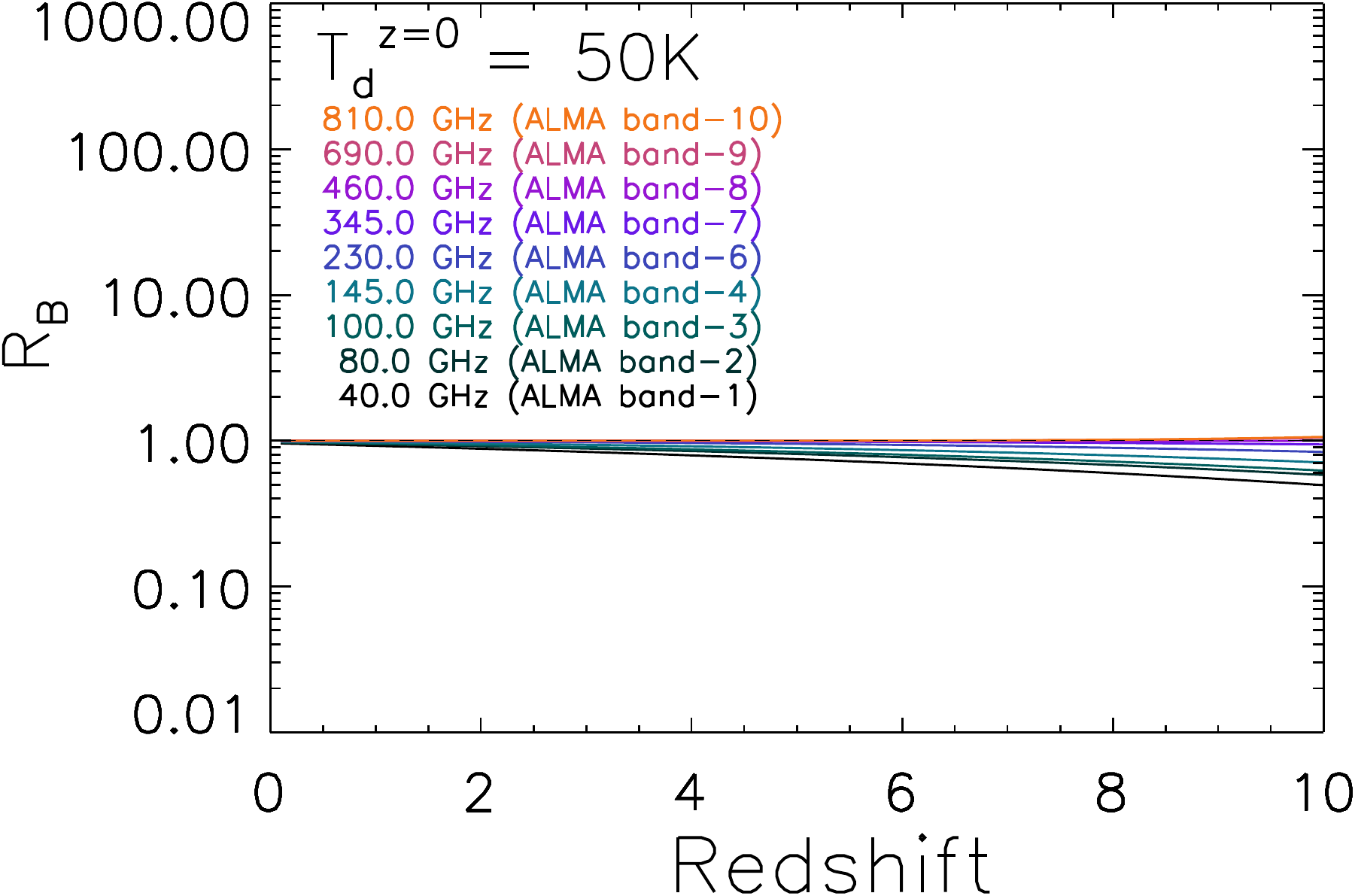}
\caption{Predicted brightness ratios ($R_{\rm B}$; Equation \ref{Rb}) between
the dust emission with the SEDs with the CMB at redshift, $z$, and with that at
redshift 0, observed in different ALMA bands ($\nu_{\rm obs}$ = 40, 80, 100,
145, 230, 345, 460, 690, and 810\,GHz). The CMB effects include both the
additional CMB heating on the dust and the background continuum subtraction.
{\it Left:} The ratios with an assumption of $T_{\rm d}^{z=0}  = 20$\,{\sc k}.
{\it Right:} The ratios with an assumption of $T_{\rm d}^{z=0}  = 50$\,{\sc
k}.} \label{comeback}
\end{center}
\end{figure*}

Fig.~\ref{comeback} presents the effects of the CMB on the dust continuum
emission by comparing the emergent flux density at $z = 0$ and that of various
redshifts. We plot $R_{\rm B}$ (Equation~\ref{Rb}) as a function of redshift
for different ALMA bands ($\nu_{\rm obs}$ = 40, 80, 100, 145, 230, 345, 460,
690, and 810\,GHz). We adopt two intrinsic dust temperatures, $T_{\rm d}^{z=0} =
20$\,{\sc k} for the quiescent cold dust on the disk and $T_{\rm d}^{z=0} =
50$\,{\sc k} for the warm dust heated by star formation or active galactic
nuclei. We find that $R_{\rm B}$ decreases with redshift at low frequencies
(dimming) and increases with redshift at high frequencies (re-brightening).
Both effects are much less pronounced in the high-temperature case ($T_{\rm
d}^{z=0}  = 50$\,{\sc k}). For $T_{\rm d}^{z=0} = 20$\,{\sc k} at $z =8$, the
brightness in ALMA band 10 is an order of magnitude higher than that at $z =
0$, while in ALMA band 3 the brightness deceases by a factor of ten compared to
that at $z = 0$. ALMA band 7 is the least affected observing frequency.

\section{Effects of the CMB on the observed brightness distributions of dust
continuum in spirals } \label{cmbdust}

In this section we use three spiral galaxies to
demonstrate the impact of the elevated CMB on the brightness distribution of
their dust continuum and thus on the recoverable morphology of their dust mass
distributions. Rather than assuming a $T_{\rm dust}$ range -- as was the past
practice for such studies \cite{DaCunha2013ApJ,Combes1999AA} -- we will
demonstrate the CMB effect on the observable brightness distribution of cold
dust continuum at high redshifts by using real $T_{\rm dust}$ maps of galaxies
in the local Universe. Such maps can now be obtained from the wealth of
IR/submm imaging data available from the {\it Spitzer Space Telescope} and the
{\it Herschel Space Observatory}. We exploit NGC\,628, M\,33, and M\,31, for
which there are high-quality $T_{\rm dust}$ maps over most of their extent
\cite{Smith2012ApJ,Galametz2012MNRAS,Xilouris2012AA} and whose levels of
star-formation activity range from the vigorous, in NGC~628 and M\,33
($\Sigma_{\rm SFR} \sim 10^{-3} - 10^{-2}$ \SFRsd\ --
\cite{Bigiel2008,Heyer2004}), to the quiescent M\,31 ($\Sigma_{\rm SFR} \sim
10^{-4} - 10^{-3}$ \SFRsd\ -- \cite{Ford2013}).

\subsection{Archival {\it Herschel} data}

To model the cold dust emission in high-redshift galaxies, we exploit
500-$\mu$m images from the {\it Herschel} science archive system. The
galaxies were observed using the Spectral and Photometric Imaging
Receiver (SPIRE -- \cite{griffin2010}). For NGC~628, the data were
obtained as part of the {\it KINGFISH} project \footnote{Key Insights
on Nearby Galaxies: \url{http://www.ast.cam.ac.uk/research/kingfish}} 
(P.I.~Kennicutt; observation ID 1342179050). The M~33 data were observed 
in the {\it HerM33es} project\footnote{Herschel M33 extended survey: \url{http://www.iram.es/IRAMES/hermesWiki/FrontPage}}
(P.I.~Kramer; observation ID 1342189079). The M~31 data were observed
in the {\it HELGA} project\footnote{The Herschel Exploitation of Local Galaxy Andromeda:
 \url{http://www.astro.cardiff.ac.uk/research/astro/egalactic/surveys/?page=HELGA}
} (P.I.~Fritz; observation ID 1342211294).

\subsection{Temperature maps of cold dust distribution}

Dust temperatures are often derived from fitting a modified blackbody.
Most galaxies, however, contain a range of dust temperatures. The warm
dust component (\Td$\sim 40-100$\,{\sc k}) often contains less mass
but contributes significantly to the fitting of single MBB, leading to
an overestimate of the cold dust temperature. Dual-MBB model fits can
better constrain the temperature for the cold dust component, which is
where the effect of the CMB plays an important role
\cite{Galametz2012MNRAS}. We simplify the fitting procedures in the
literature \cite{Smith2012ApJ,Xilouris2012AA,Galametz2012MNRAS} and
repeat the fit to derive cold dust temperature maps for M~31, M~33,
and NGC~628, by fixing the emissivity indices for both cold and warm
components. Here we do not consider any variation of $\beta$ within
the galaxies, to minimise the degeneracy between the dust temperature
and $\beta$. The adopted value, $\beta = 2$, is often used for
modelling high-redshift galaxies globally. This assumption allows us
to see how the dust temperature is changed solely by the CMB effect.
We obtain a robust determination of the cold dust temperature
distribution based on maps at multiple wavelengths. We exclude pixels
below the 3$\sigma$ detection limit in all bands.

From the $T_{\rm dust}^{z=0} $ maps we obtain $T_{\rm dust}^{z}$ maps
at a given redshift $z$, accounting for the CMB dust heating
\cite{Papadopoulos2000ApJ,DaCunha2013ApJ,Combes1999AA}. In order to
show solely the effect of the CMB on the emergent dust continuum
brightness distribution in the source rest frame we do not apply the
cosmological correction $I_{\nu _{\rm obs}(r)}=(\nu_{\rm obs}/ \nu
_{\rm em})^3$\, $I_{\nu_{\rm em}}(r)$, where $ I_{\nu _{\rm em}(r)}$
is the source rest-frame brightness distribution. Doing so would
introduce another $\rm (1+z)^{-3}$ dimming that could make it even
harder to discern the CMB-immersed brightness features. Finally, to
obtain the brightness images at various observing frequencies, we scale the {\it
Herschel} 500-$\mu$m images by the factor $R^{\prime}_{\rm B}$
(Equation~\ref{aim}) at both redshift $z$ and $z = 0$.

\begin{figure}
 \begin{center}
\includegraphics[scale=.60]{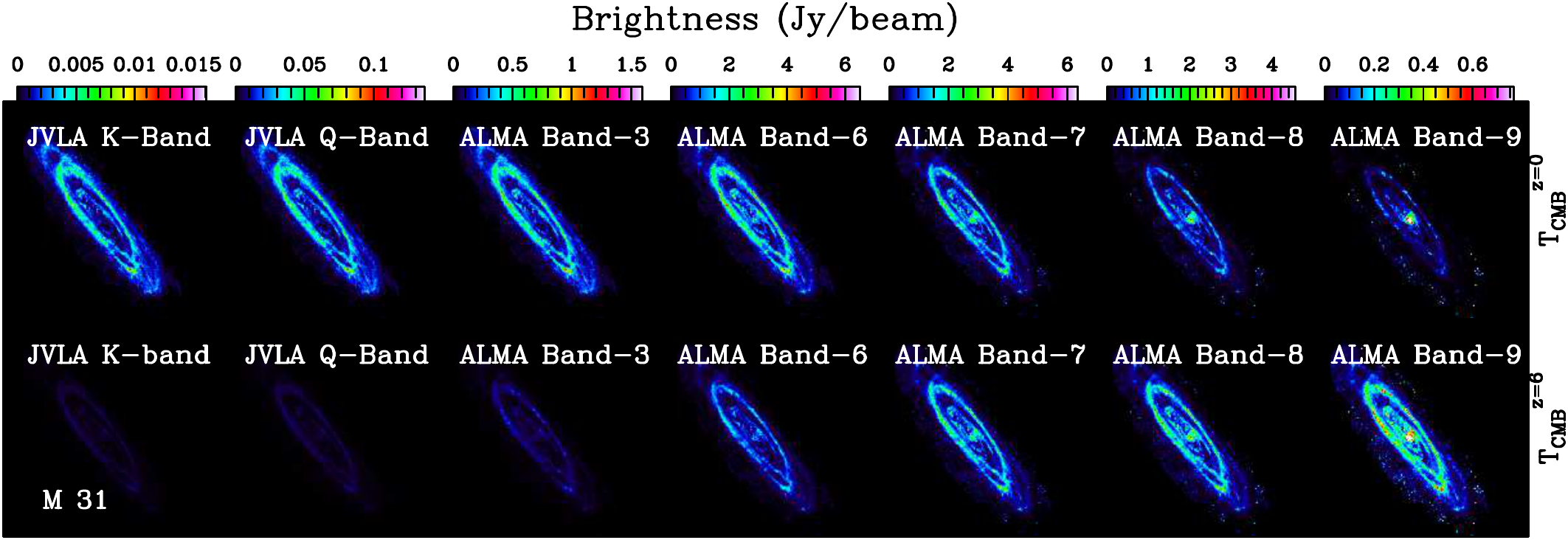}
\includegraphics[scale=.60]{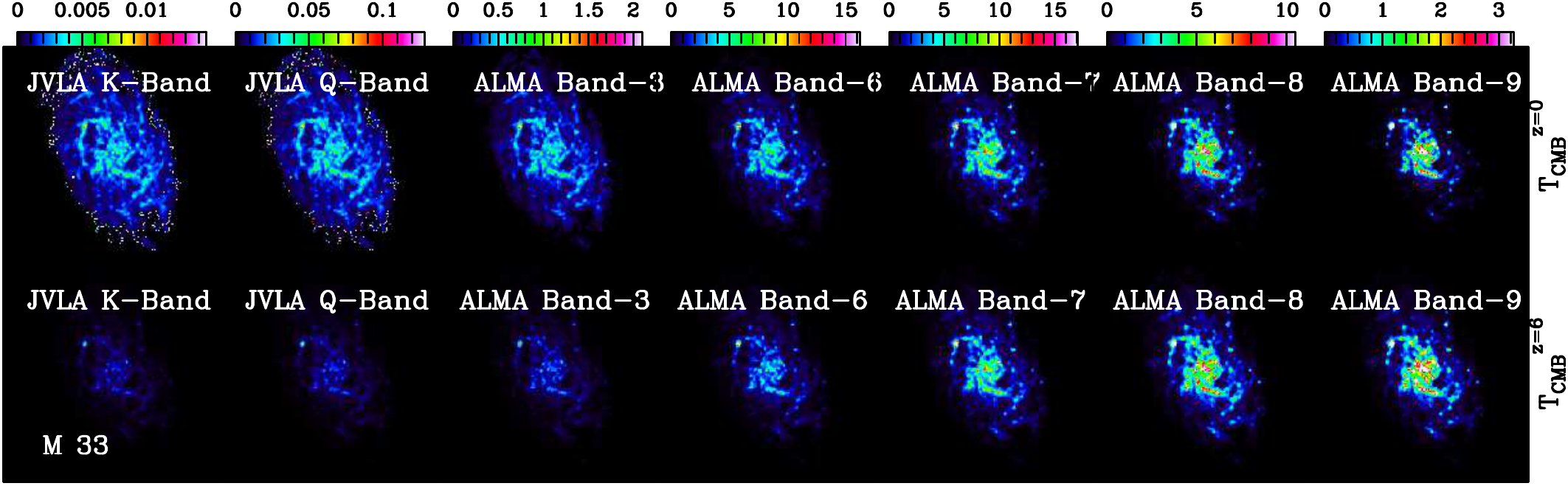}
\includegraphics[scale=.60]{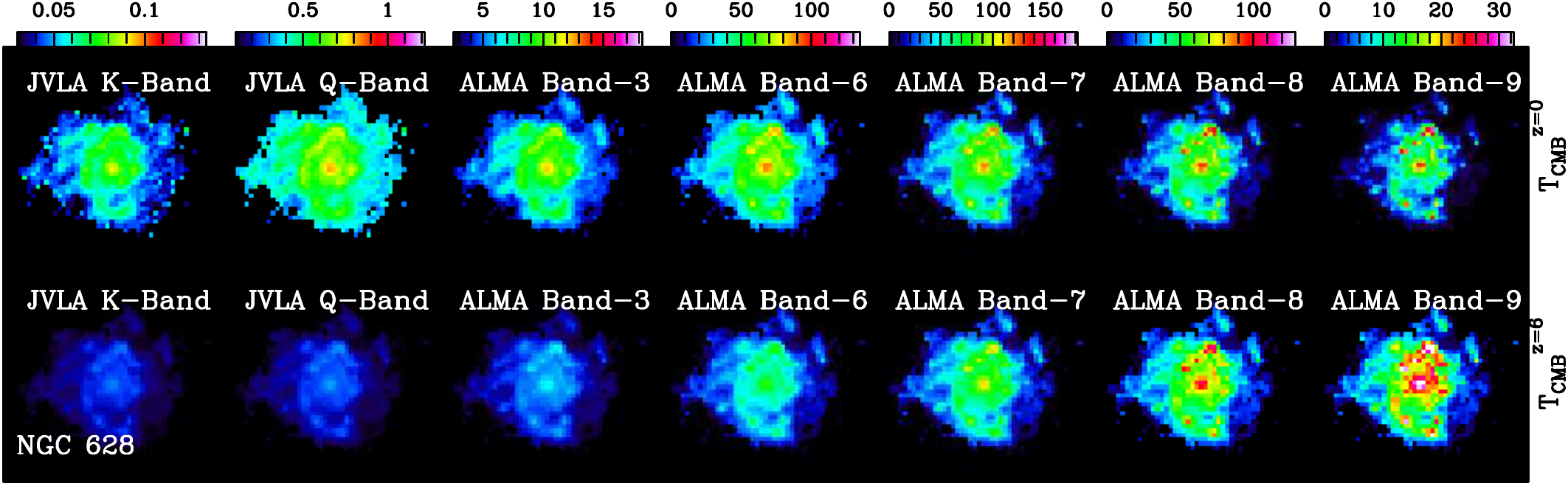}
\caption{Simulated images of the cold dust continuum emission of M\,31, M\,33
 and NGC\,628 at $z = 6$, observed in different bands (25, 45,
 100, 230, 345, 460, 690\,GHz) of the JVLA and ALMA. The images are scaled
 from 500-$\mu$m {\it Herschel} images at $z =0$ with the $R^{\prime}_{\rm
 B}$ factors, and are displayed in the source rest frame. For each galaxy
 we plot the continuum emission with the CMB temperatures of $z = 0$
 ($T^{z=0}_{\rm cmb} = 2.725$\,{\sc k}; upper panels) and $z = 6$ ($T^{z=6}_{\rm cmb} =
 19.075$\,{\sc k}; lower panels). }
\label{dustmaps}
\vspace*{-15pt}
\end{center}
\end{figure}

\begin{equation} \label{aim}
        R^{\prime}_{\rm B} = ( \frac{\nu}{\nu_{\rm 500 \mu m}})^{\beta + 2}
 \frac { J[T_{\rm d}(z), \nu] - J[T_{\rm CMB}(z), \nu] }
 { J[T_{\rm d}^{z=0} , \nu_{\rm 500 \mu m}] - J[T_{\rm CMB}^{z=0}, \nu_{\rm 500 \mu m}] }
\end{equation}

\noindent
where $J(T, \nu)=h\nu/k_{\rm B}/[\exp(h\nu/k_{\rm B}T)-1]$ is the
Planck radiation temperature at frequency, $\nu$, with a dust
temperature, $T_{\rm dust}$. And $\beta$ is the dust emissivity index. We
fix $\beta =2$.  Varying $\beta$ does not change our conclusions.

\begin{figure}
 \begin{center}
\includegraphics[scale=.40]{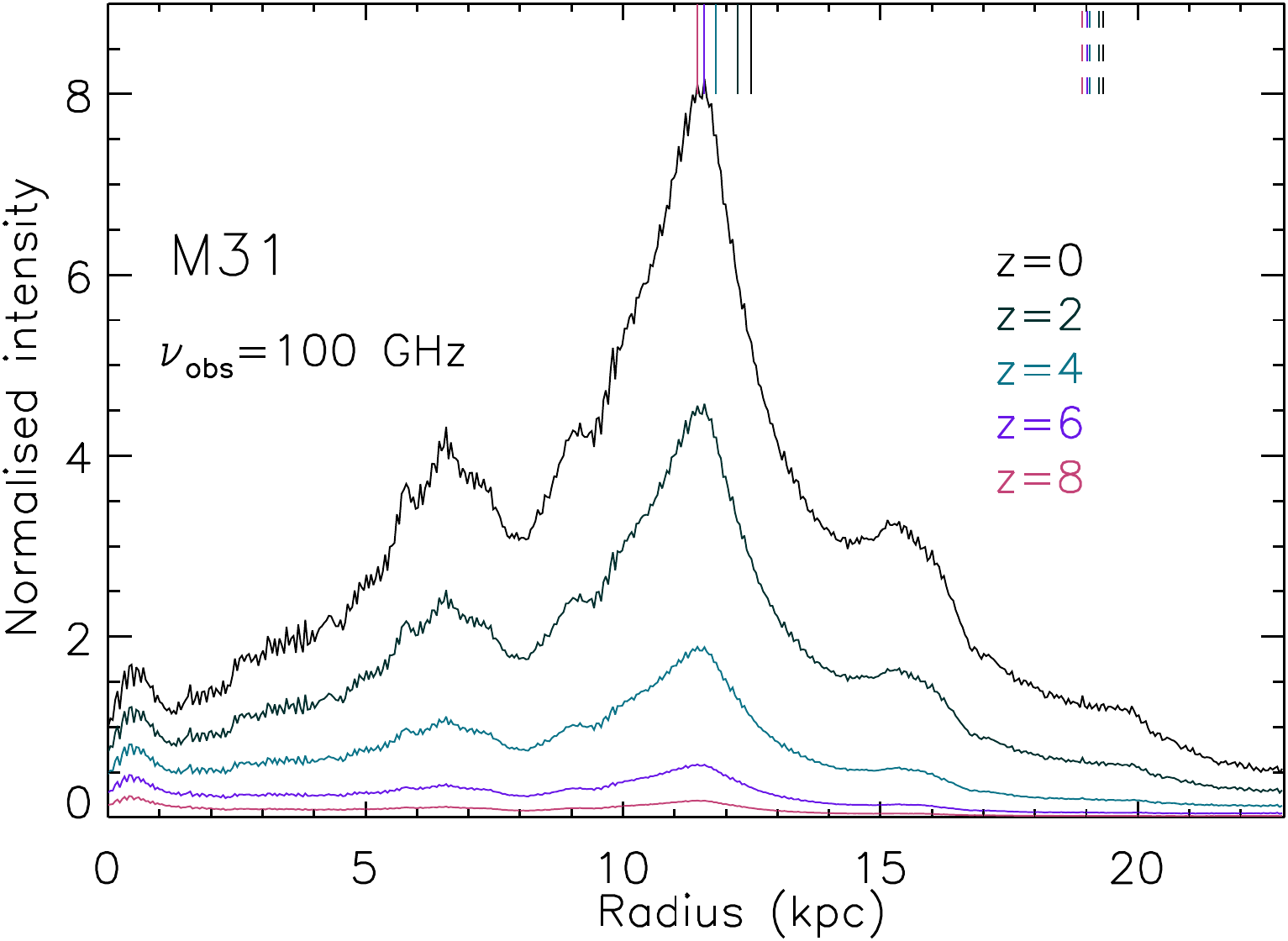}
\includegraphics[scale=.40]{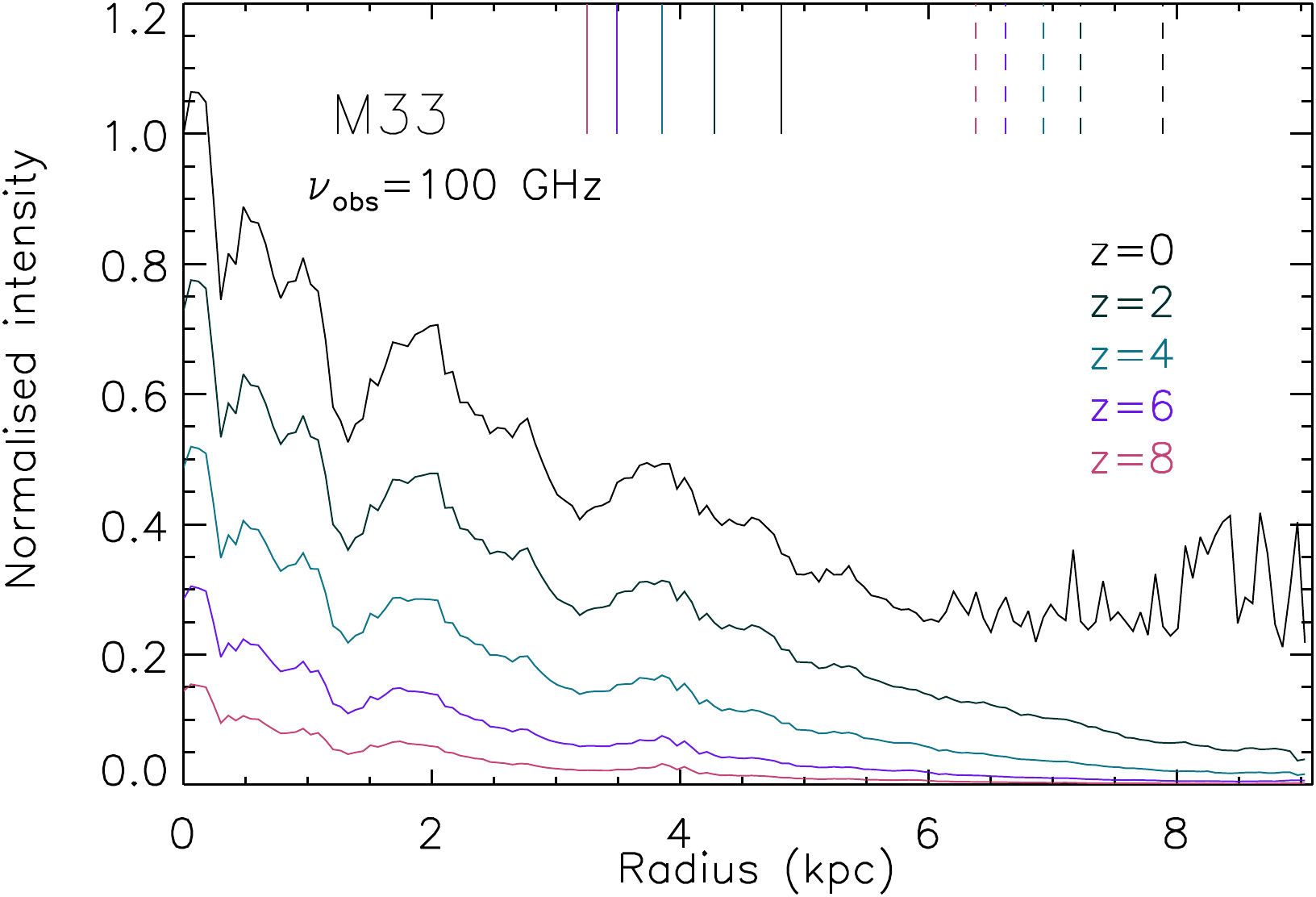}
\includegraphics[scale=.40]{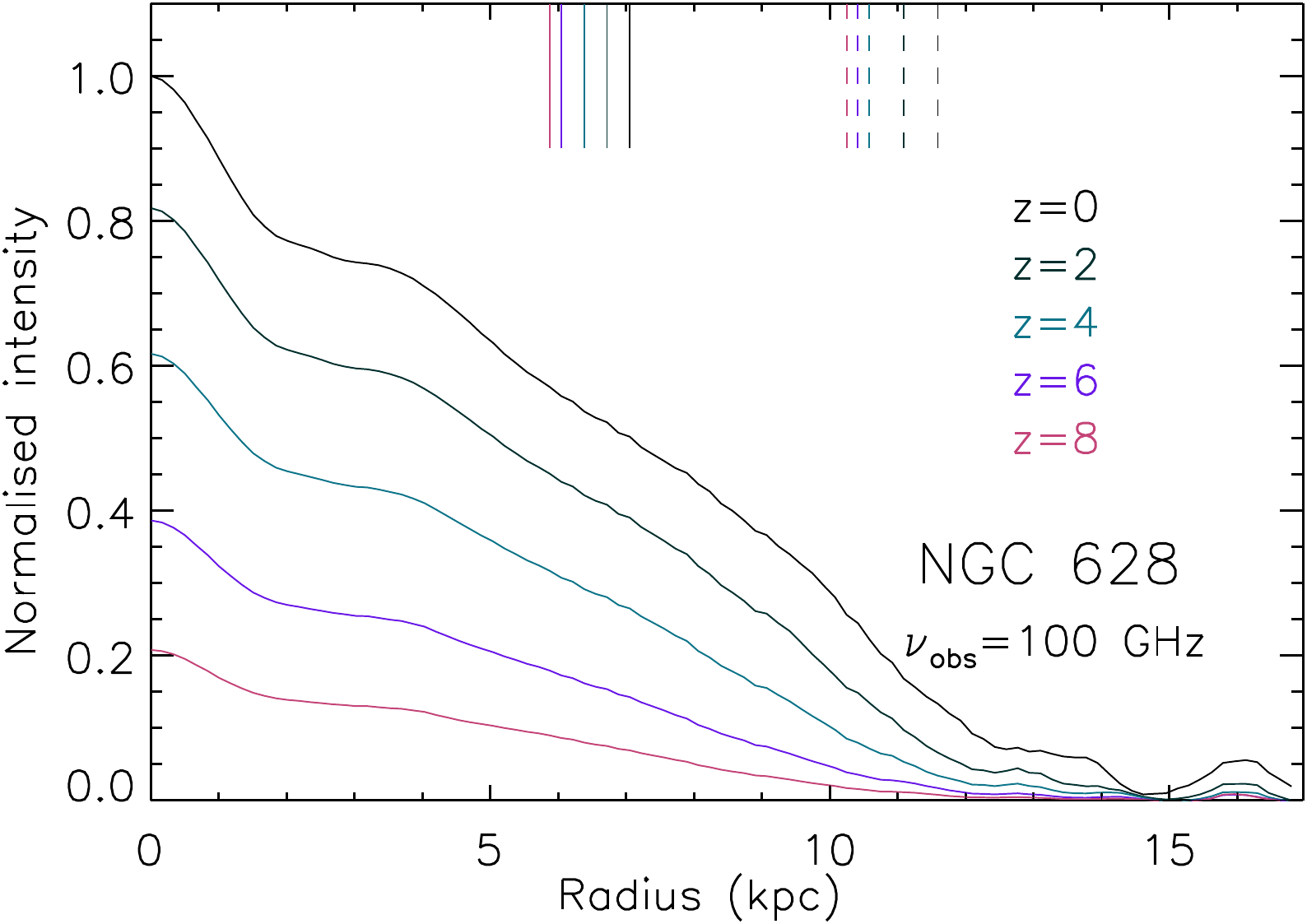}
\caption{Radial distributions of the dust continuum emission of M\,31, M\,33
 and NGC\,628 emitting at an observing frequency of 100\,GHz at redshifts 0, 2,
 4, 6, and 8, respectively, shown in different colours.  We make concentric
 elliptical rings with the inclination and position angles of these galaxies,
 and obtain the average flux in each ring. The radial distribution plots are
 then normalised to emission at the central positions of the galaxies at $z
 = 0$. The vertical solid lines show the half-light radii ($R_{50}$), and the
 vertical dashed lines show the 90\% light radii ($R_{90}$).}
 \label{radial_dust}
%\vspace*{-15pt}
\end{center}
\end{figure}

We compare the re-scaled maps for the CMB at $z= 0$ and $z=6$ as an
example. The re-scaled images are shown in Fig.~\ref{dustmaps} and they
correspond to the dust emission brightness distributions at rest frequency $\nu$
{\it in the source rest frame}, i.e.\ what an observer at the given redshift
would see. This isolates the effect of the enhanced CMB on the observed
brightness distribution from all the other factors that also affect it, e.g.\
cosmological size changes, $(1+z)^{-3}$ dimming, telescope sensitivity, etc. We
then examine the relative changes before and after considering the CMB effects
in these galaxies. This simple re-scaling assumes that: a) the dust emission is
optically thin, and b) dust optical depths do not have any dependence on
$T_{\rm dust}^{z}$. These hold for the dust emission in local galactic disks.

From Fig.~\ref{dustmaps}, the CMB effect on the brightness distributions of the
observed dust emission is obvious, with the contrast between the cold dust
emission and the CMB gradually disappearing at relatively lower frequencies.
Besides the `shrinking' of the observed sizes of the dust emission in all three
galaxies (expected since the cold dust phase normally traces the most extended
regions of galactic disks), we also see secondary effects where the warm/cold
dust emission contrast changes {\it within} the observed disks. We note
that for ALMA band 7 observations the global scale lengths and the dust
structures are only modestly affected by the CMB, since it lies in the
transition zone between the Rayleigh-Jeans and the Wien regimes for the cold
dust of these galaxies (see \S\ref{math}).

In Fig.~\ref{radial_dust}, we plot the radial distributions of the dust
continuum emission at the observing frequency of 100\,GHz at $z = 0$, 2, 4, 6
and 8, respectively. The colours show the redshifts. We make concentric
elliptical rings with the inclination and position angles of these galaxies,
and obtain the average flux within each ring. The radial distribution plots are
then normalised to the flux in the central positions of galaxies at $z = 0$.
The vertical solid lines show the half light radii ($R_{50}$), within which
half of the total flux of the galaxy is found. The vertical dashed lines show
the 90\% light radii ($R_{90}$).

The effect of the CMB on the observed spatial distributions of dust masses is
important since such morphological characteristics are often used to infer
whether disk instabilities are driving star formation in disks. In the case of
the globally cold M\,31, the effect of a rising CMB is most dramatic, with the
entire galaxy `fading' into the CMB. It is worthwhile to point out that the
$R_{50}$ and $R_{90}$ radii are measured without taking noise into account, so
they only show the relative change between the central region and disk when we
have unlimited sensitivity. The effect of the CMB on the observed dust
continuum distributions and scale lengths is obvious.  Here we must also note
that the cold dust component will be present whenever dust lies far from
radiation sources (stars or active galactic nuclei) that could warm it, like
the dust reservoirs expected in the gas outflows or inflows found in the local
and the distant Universe \cite{Cicone2014AA,Cicone2015AA,George2014MNRAS}. Thus
the effect of the rising CMB on the brightness distributions of dust emission
will not be confined solely to galactic disks.

\section{Effects of the CMB on the observed morphology and kinematics of the molecular gas}
\label{COmodel}

In what follows we investigate the impact of the CMB on molecular gas mass
tracers, namely the CO and C\,{\sc i} emission lines, and on the H$_2$ gas
velocity fields marked by CO and C\,{\sc i}. We select the NGC\,628 galaxy, for
which a large and fully sampled $J$=2--1 map is available \cite{Leroy2009AJ}.
We consider the lowest $J$ transitions of CO, i.e.\ the $J$=1--0, 2--1 and 3--2
lines. The first two trace the global H$_2$ gas in galaxies, irrespective of
the thermal state or density (as long as $n\gtrsim 10^3$\,cm$^{-3}$, which is
the case for most CO-rich H$_2$ gas in galaxies), and are thus the spectral
lines of choice when unbiased views of H$_2$ gas mass and H$_2$ velocity fields
are sought. CO $J$=3--2 is the last CO transition to have non-negligible
contributions from the cold, low-density, non-SF gas, though most of its
luminosity comes from the warm and dense H$_2$ gas found near SF sites. Any
higher-$J$ transitions are unsuitable as global tracers of H$_2$ gas mass and
velocity fields. We also compute the CMB effects on the two neutral atomic
carbon lines, C\,{\sc i} 1--0 and 2--1, which have been shown to be also good
tracers of H$_2$ gas mass and galactic dynamic mass, potentially better than
the low-$J$ CO transitions\cite{Papadopoulos2004MNRAS,Bisbas2015,Zhang2014}.
C{\sc ii} is an excellent tracer of gas dynamics and is barely affected by the
CMB, however the C{\sc ii} emission mostly arises from both photon dominated
regions and H{\sc ii} regions, making it not solely sensitive to the H$_2$ gas
\cite{2014ApJ...784...99G}.

We do not have a multiplicity of CO lines per position within NGC\,628 that
would allow us to determine $n_{\rm H_2}$, $T_{\rm kin}$ and $d v/d r$ (the
average gas velocity gradient, which depends on the dynamical state of the
H$_2$ gas). These parameters, once determined, could allow the computation of
the emergent CO line intensities under different CMB backgrounds using a
typical large velocity gradient (LVG) radiative transfer approach.  However,
the lack of spatially resolved CO SLEDs necessitates a more conservative
approach where we set: kinetic temperature $T_{\rm kin} = T_{\rm d}$, H$_2$
number density $n_{\rm H_2}$=10$^3$ cm$^{-3}$ and velocity gradient $dv/dr$= 1
\kmspc, which are typical for quiescent non-SF H$_2$ clouds in local spiral
galaxies. We use the available $T_{\rm d}^{z}$ maps as input to a standard
LVG code to compute the emergent CO line brightnesses scaled from the available
local CO $J$=2--1 map.  Unlike the case for dust emission and its optical
depths, the optical depths of CO lines do depend on $T_{\rm kin}^{z}$ and thus
do not cancel out.

\subsection{Archival CO data}

We adopt the existing fully sampled CO $J$=2--1 data cube of NGC~628 in the
HERA CO-Line Extragalactic Survey (HERACLES) archive\footnote{\url{http://www.mpia.de/HERACLES}}. The data reduction has been
described in \cite{Leroy2009AJ}. The data cube has been converted to the main
beam temperature scale, which represents the brightness temperature of CO
lines. The angular resolution is $\sim13''$, corresponding to a linear scale of
$\sim 0.6$\,kpc.

\subsection{Temperature maps of CO}\label{COtemperature}

In order to match the high angular resolution CO data of NGC~628, we convolve
all the data to an angular resolution of $18''$ (the resolution of the {\it
Herschel} 250-$\mu$m image), and generate a \Td\ map without the 350- and
500-$\mu$m data. The new \Td\ map is similar to the previous low-resolution \Td\ map
adopted in modelling the dust emission (including the 350- and 500-$\mu$m data;
see \S\ref{cmbdust}), with $<$15\% difference in \Td. 

\subsection{Radiative transfer modelling with LVG}

In order to model the excitation conditions of molecular gas, which
may not be in a local thermal equilibrium, the radiative transfer should
be solved simultaneously for multiple population levels. We adopt a
commonly used LVG assumption \cite{Scoville1974,Scoville1974} wherein
the line emission affected via self absorption or induced emission
only occurs in a local region, and the size is of the order of the
local velocity dispersion (thermal and micro-turbulent) divided by a
velocity gradient, \dvdr.

\begin{figure}
 \begin{center}
\includegraphics[scale=.45]{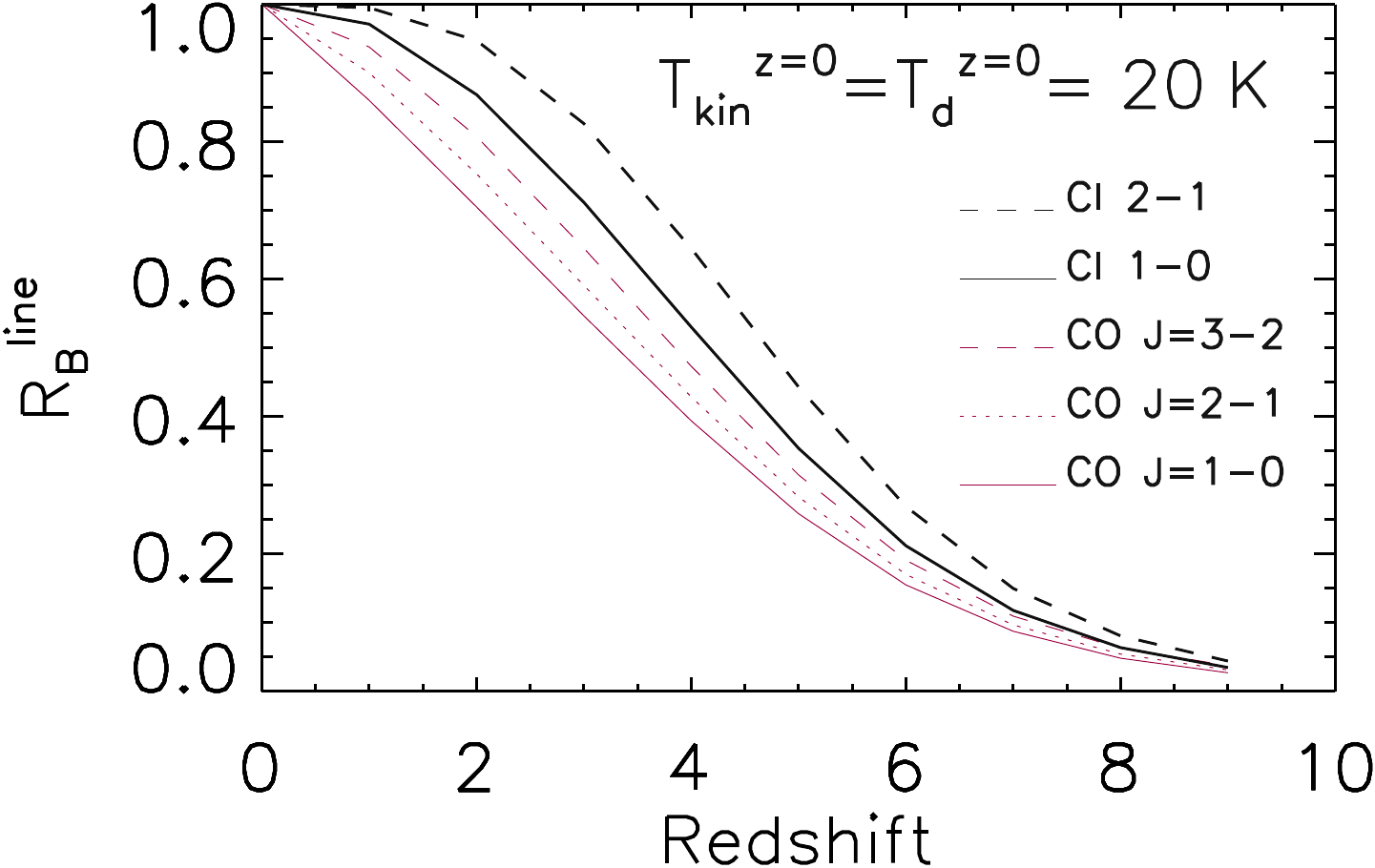}
\includegraphics[scale=.45]{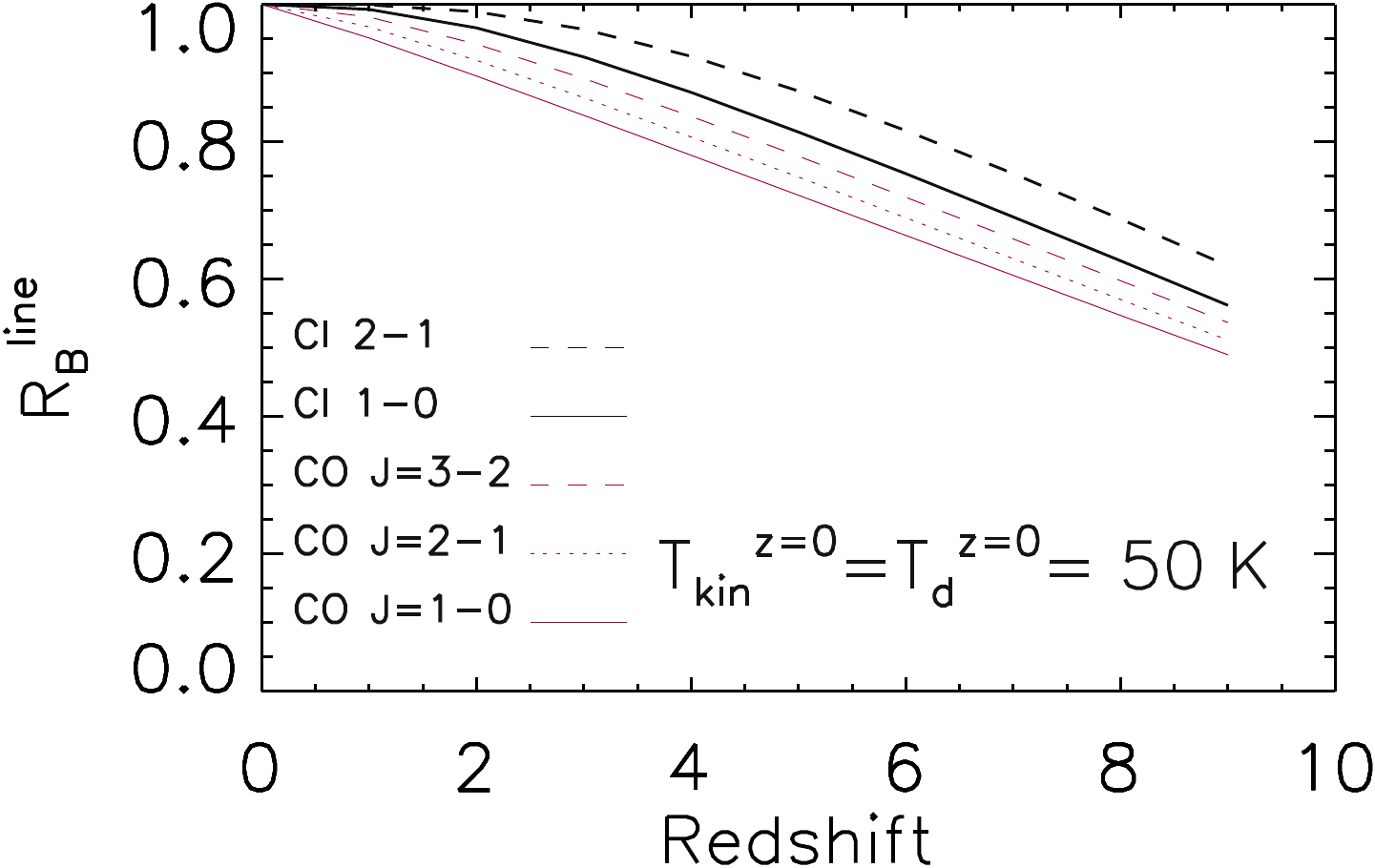}
\caption{Predicted brightness ratios of line emission ($R_{\rm B} ^{\rm
line}$) between at redshift $z$ and at $z=0$. The ratios are obtained
using LVG modelling which assumes a molecular density of 10$^3$ cm$^{-3}$ and
virialised conditions. {\it Left:} $R_{\rm B} ^{\rm line}$ when assuming
$T_{\rm kin}^{z=0}$ = $T_{\rm d}^{z=0}$ = 20\,{\sc k}. {\it Right:} $R_{\rm B}
^{\rm line}$ when assuming $T_{\rm kin}^{ z=0}$ = $T_{\rm d}^{z=0}$ = 50\,{\sc
k}.}
\label{Rsline}
\end{center}
\end{figure}

We used the LVG code, {\sc myradex}\footnote{{\sc Myradex} adopts the
same equations as RADEX, and has no convergence problems, which
appear in RADEX in certain parameter space:
\url{https://github.com/fjdu/myRadex}}, to model the CO ladders and determine
the predicted CO line intensities. We adopt the geometry of a uniform radially
expanding sphere, which has an escape probability of $(1 - \exp(-\tau))/{\tau}$, where $\tau$ is the optical depth of a given transition. We
use the molecular collisional rates from the Leiden Atomic and Molecular
Database (LAMDA)\footnote{\url{http://home.strw.leidenuniv.nl/~moldata}}. The
relative abundances to H$_2$ are $8\times 10^{-5}$ for CO and $5\times 10^{-5}$
for C\,{\sc i}, respectively \cite{Weiss2005}. We adopt a H$_2$ number density
of 10$^3$ cm$^{-3}$, which is a representative value for the bulk of molecular
gas. We further assume the molecular gas is in a virialised condition, and set
\dvdr\ to be 1\,\kms, since typically quiescent non-SF H$_2$ clouds in galaxies
are self-gravitating. The CMB temperature, $T_{\rm CMB}^{z}$, varies
with redshift as $T^{z}_{\rm CMB}= T_{\rm CMB}^{z=0} \times (1+z)$, where $T_{\rm
CMB}^{z=0} $ is 2.73\,{\sc k}. Using these inputs at $z = 0$, we obtain the
$R_{\rm B}^{\rm line}$ factor at various temperatures and redshifts.

\begin{figure}
 \begin{center}
\includegraphics[scale=.80]{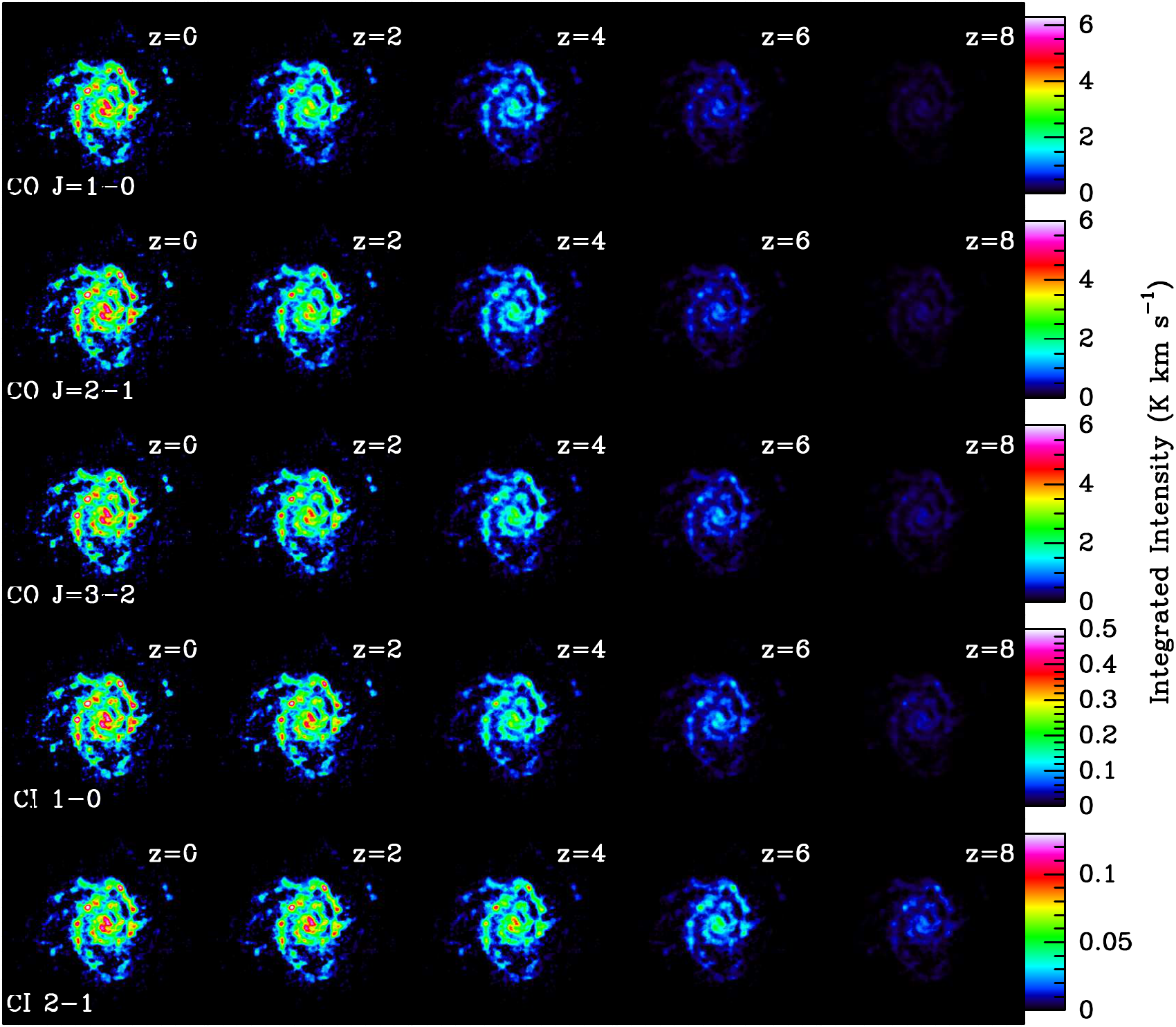}
\caption{Simulated velocity-integrated brightness temperature
  (moment-0) maps of the CO and C\,{\sc i} lines of NGC\,628, at
  different redshifts.  The images are displayed in the source rest
  frame with units of \Kkms. From top to bottom: CO $J$ = 1--0, CO $J$
  = 2--1, CO $J$ = 3--2, C\,{\sc i} 1--0, and C\,{\sc i} 2--1. We
  assume that $T_{\rm kin} = T_{\rm d}$ and that the H$_2$ gas is
  virialised with a uniform number density of 10$^3$\,cm$^{-3}$.}
\label{CO_CI_maps}
\end{center}
\end{figure}

\begin{equation} \label{line}
        R_{\rm B} ^{\rm line} = \frac { J[T^{\rm line}_{\rm ex}(z), \nu^{\rm line}_{\rm rest}] - J[T_{\rm CMB}(z), \nu^{\rm line}_{\rm rest}] }
        { J[T^{\rm CO\, J=2-1} _{\rm ex}(0), \nu^{\rm CO\, J=2-1}_{\rm rest}] - J[T_{\rm CMB}(0), \nu^{\rm CO\, J=2-1}_{\rm rest}] }
 [ \frac { 1- e ^{- \tau _{\rm line} (z) } }
 { 1- e ^{- \tau _{\rm CO\, J=2-1} (0) } }]
\end{equation}

\noindent
As in the case for the dust continuum emission, we have the (source)-(CMB)
brightness terms, but the optical depth terms for the line emission now do not
cancel out. In Fig.\ref{Rsline}, we investigate the effects of the CMB on the H$_2$ gas tracers by comparing their brightness at local  
and at various redshifts.
We plot $R_{\rm B} ^{\rm line}$ as a function of redshift for different
transitions of CO and C\,{\sc i}. Then we apply Equation~\ref{line} to the $z =
0$ CO $J$=2--1 map in order to scale it to other CO transitions and to the two
transitions of C\,{\sc i} at different redshifts.

\begin{figure}
 \begin{center}
 \includegraphics[scale=.40]{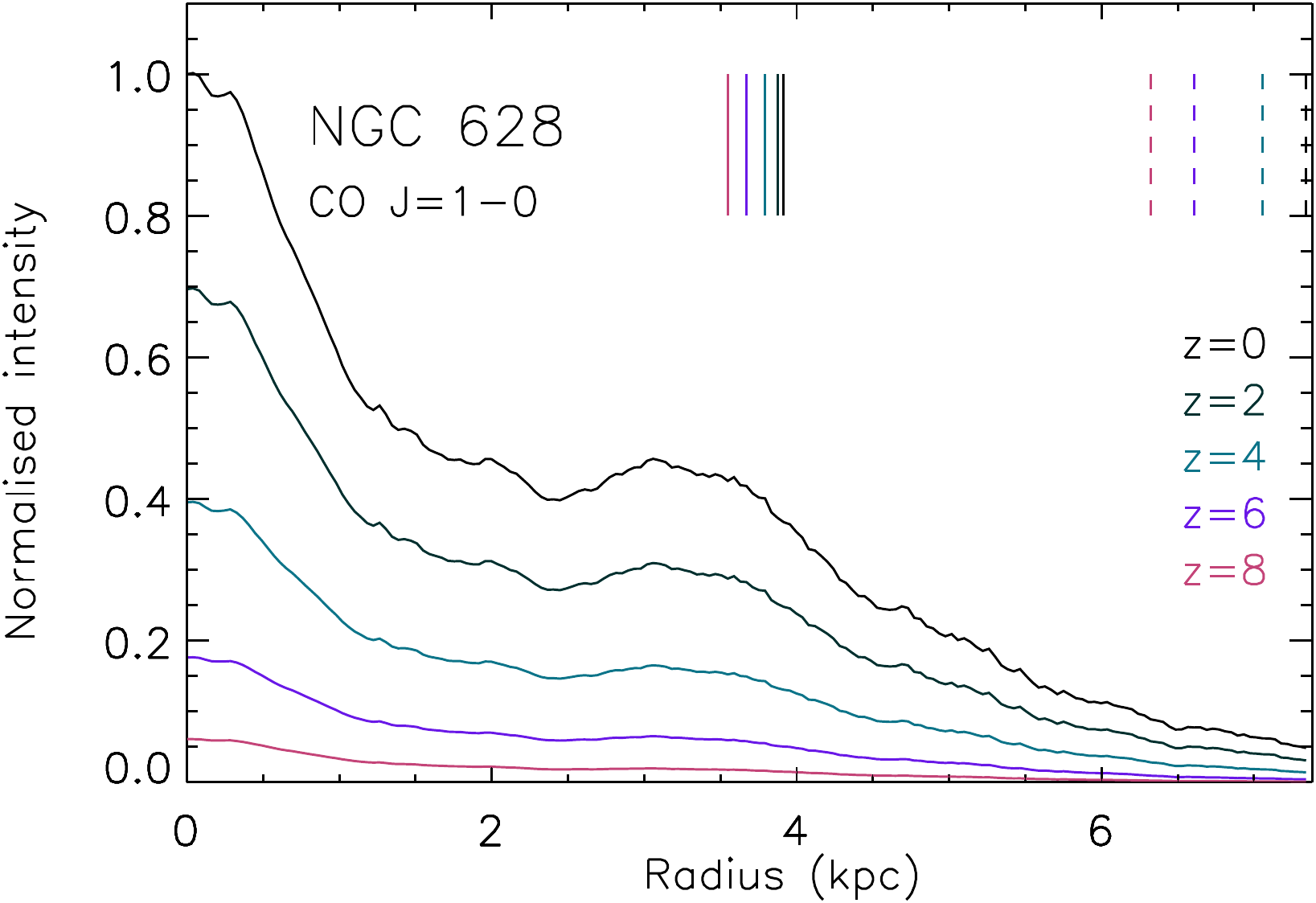}
 \includegraphics[scale=.40]{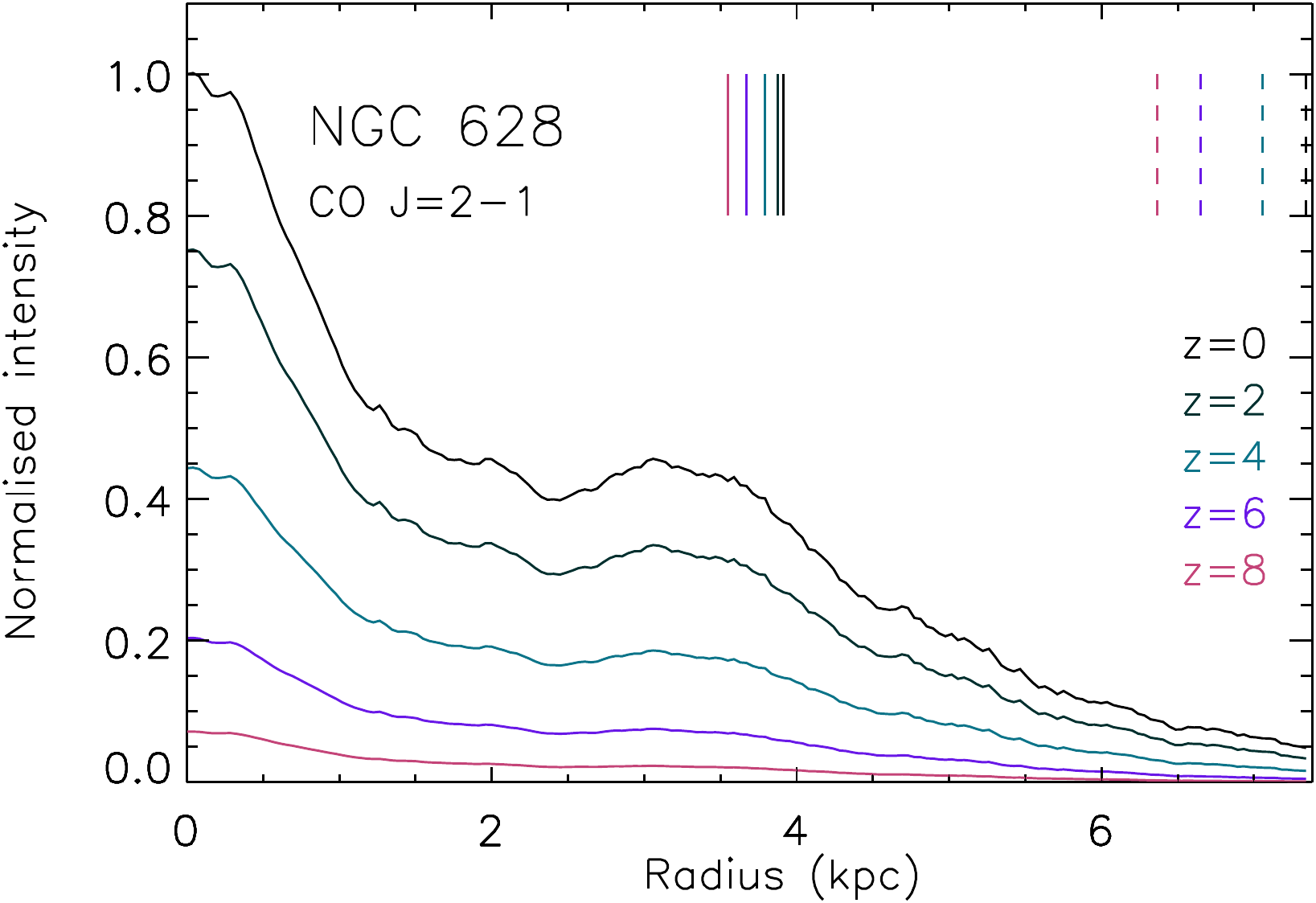}
 \includegraphics[scale=.40]{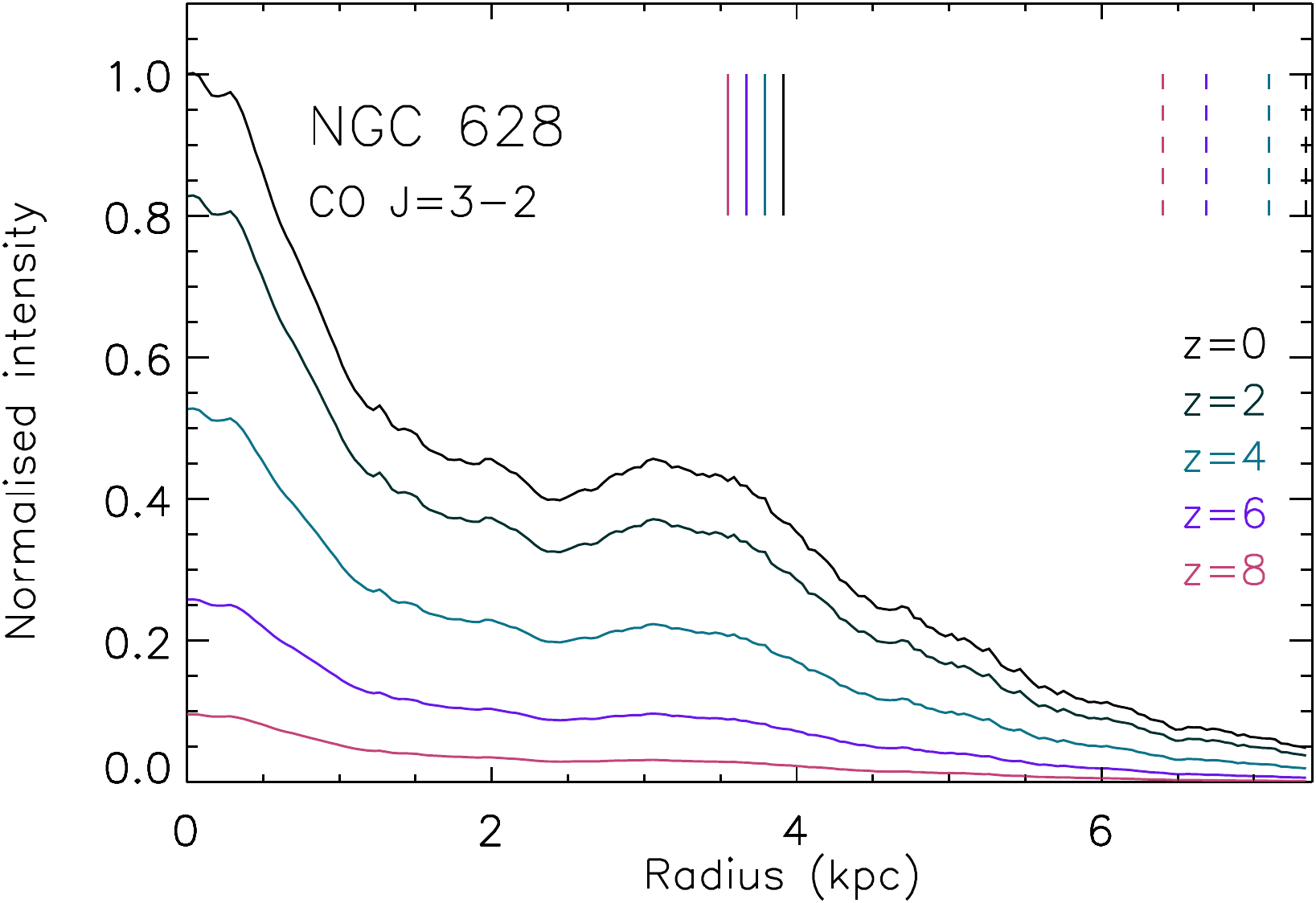}
 \includegraphics[scale=.40]{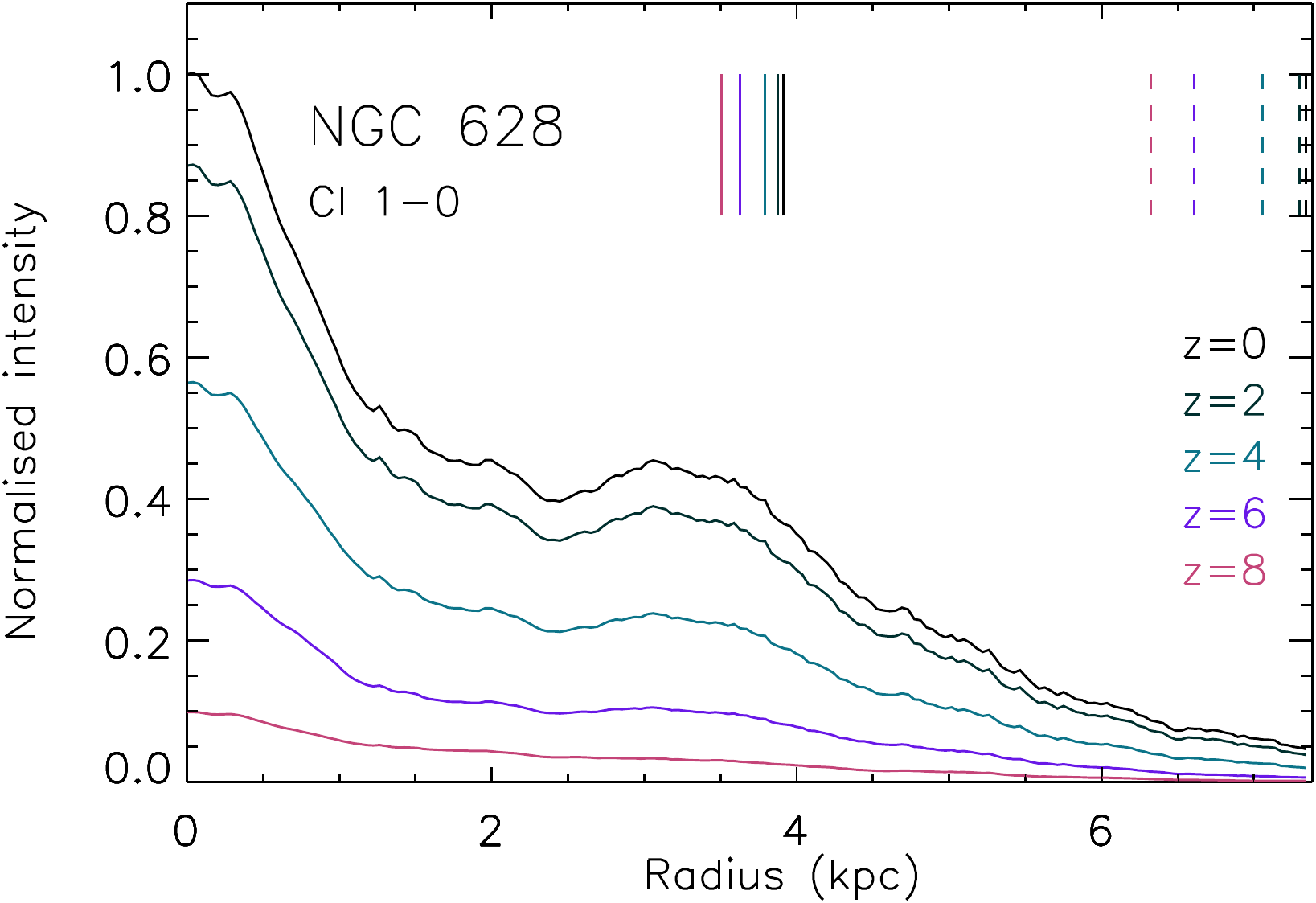}
 \includegraphics[scale=.40]{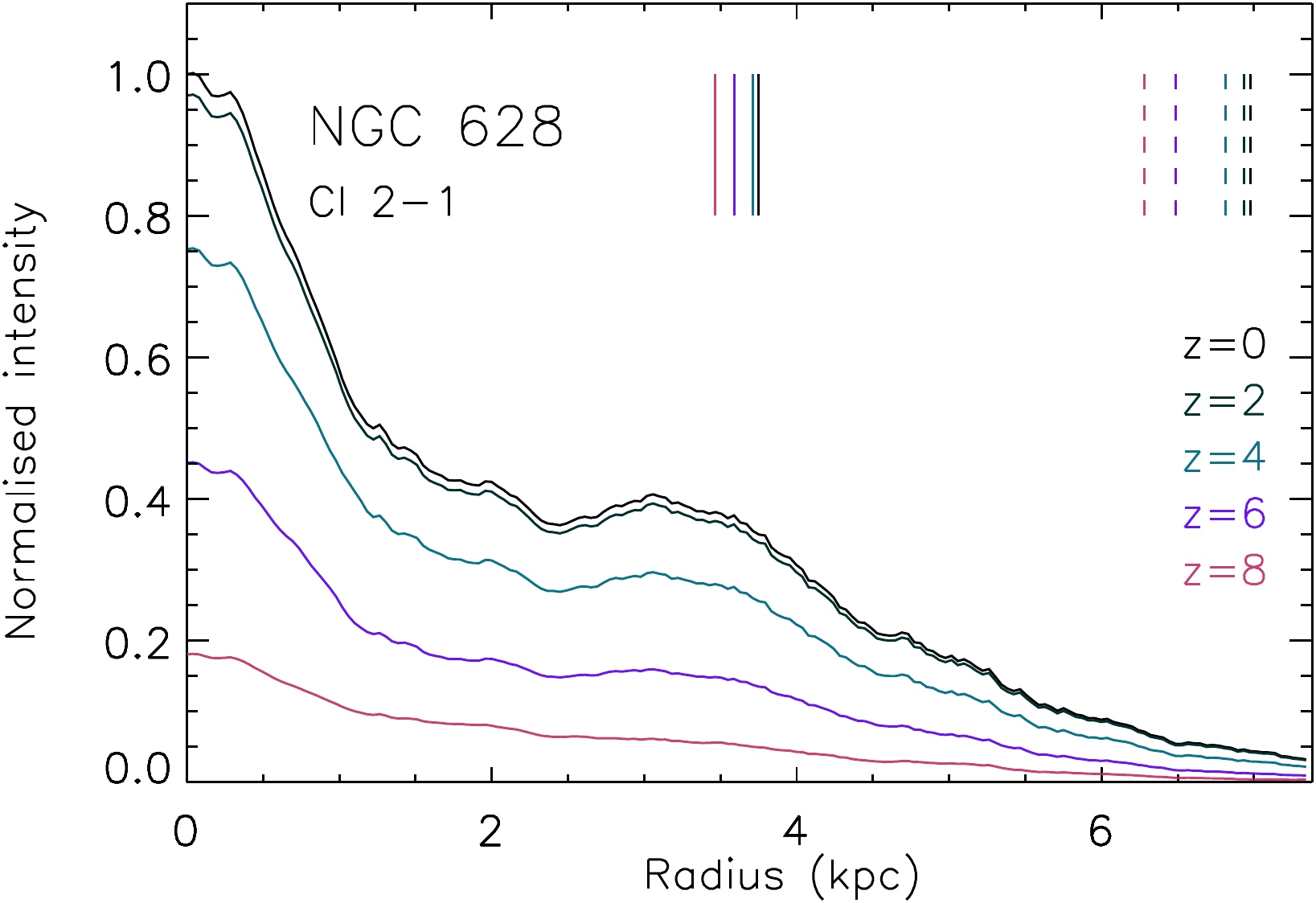}
\caption{Radial distributions of the simulated CO $J$ = 1 -- 0, CO $J$ =
 2 -- 1, CO $J$ = 3 -- 2, C\,{\sc i} 1 -- 0, and C\,{\sc i} 2 -- 1
 emission of NGC~628 at redshift 0, 2, 4, 6, and 8, respectively. 
 The colours show the redshifts. Similar to Fig.~\ref{radial_dust}, we
 measure the average flux in each concentric elliptical rings, and 
 normalise it with the flux in the central position at $z = 0$. 
 The vertical solid lines show the half-light radii 
 ($R_{50}$), and the vertical dashed lines show the 90\% light radii
 ($R_{90}$).}
\label{radial_gas}
\end{center}
\end{figure}

In Fig.~\ref{CO_CI_maps} we show the velocity-integrated (over the entire
velocity range) brightness temperature distribution of CO $J$=1--0, 2--1, 3--2,
and C\,{\sc i} 1--0, 2--1 lines in the source rest frame, for different
redshifts. The effect of the CMB is again obvious, with the contrast of the CO
lines from the cold molecular gas distributions diminishing for redshifts $z\ge
2$. We note that our conservative approach will artificially suppress the
emergent CO $J$=1--0, 2--1, 3--2 line brightnesses from the warm/dense SF
regions of NGC\,628. This is because $T_{\rm kin}$ is often significantly
higher than $T_{\rm dust}$ for the H$_2$ gas in SF regions, and $dv/dr$ is also
much larger than the quiescent value adopted. Thus the actual contrast observed
between the CO line brightness from the warm H$_2$, and the CMB-affected cold
H$_2$ gas regions at high redshifts, will be even higher than that shown in
Fig.~\ref{CO_CI_maps}. This then makes CMB-affected CO line brightness
distributions even more biased, with our simulation presenting only the minimum
effect.  In a real galaxy with clumpy star formation regions, the underlying
H$_2$ mass distribution will be even harder to recover, with only the
warm/dense `spots' detected while the colder, CMB-affected regions may be not
detectable. {\it Such CMB-affected images of CO lines would imply an underlying
H$_2$ distribution clumpier than it actually is.}

\begin{figure}
 \begin{center}
 \includegraphics[scale=.80]{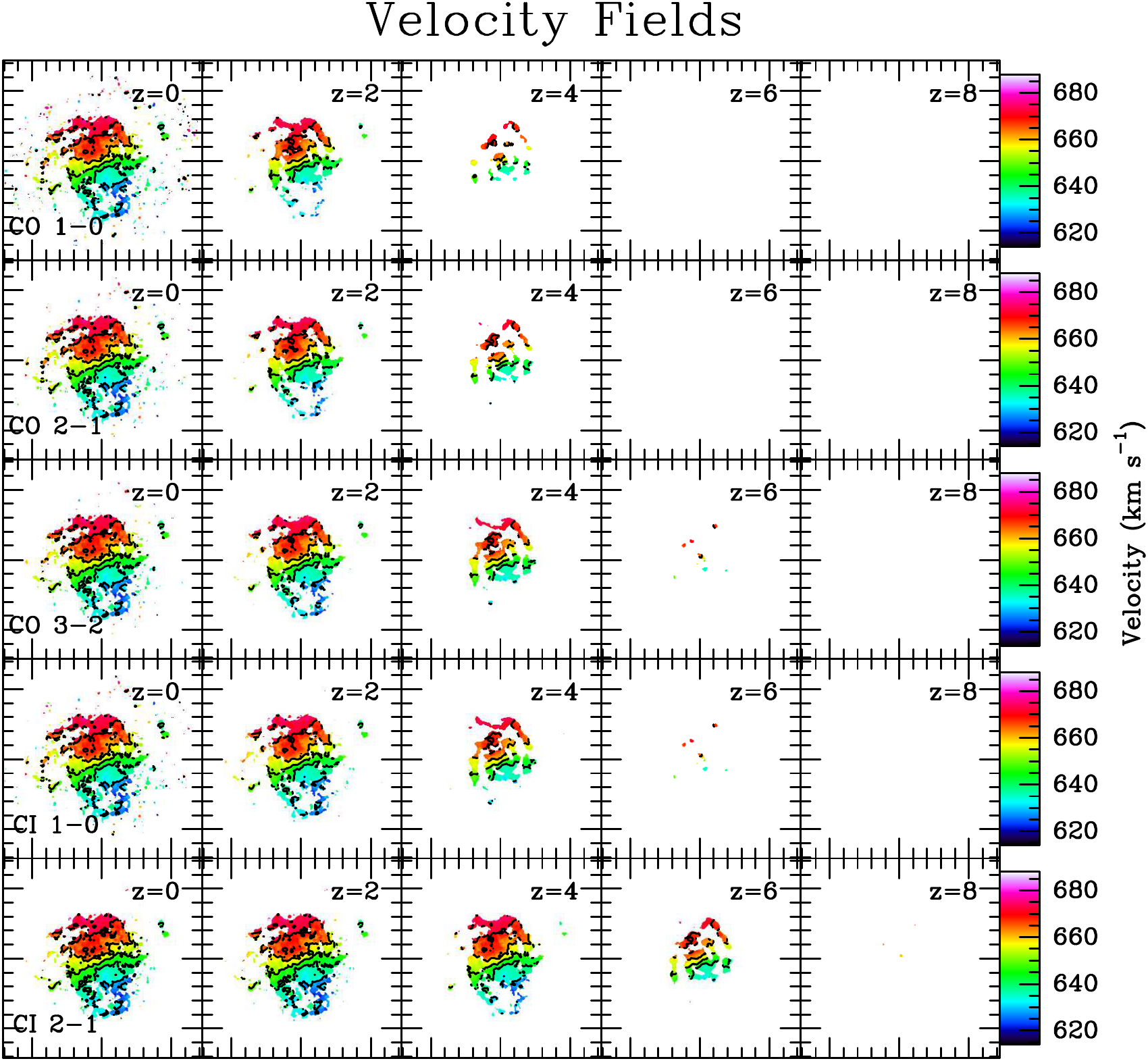}
\caption{Simulated velocity field (moment-1) maps of CO $J$ = 1 -- 0, CO $J$ =
 2 -- 1, CO $J$ = 3 -- 2, C\,{\sc i} 1 -- 0, and C\,{\sc i} 2 -- 1
 emission of NGC~628 at different redshifts. The images are displayed in the
 source rest frame. The observing frequency varies with redshift and
 transition. The contours are from 630 to 690 \kms\ with steps of 10
 \kms. We use a cutoff of $I_{\rm line} > 3 \sigma$ in generating all
the velocity-field maps. } \label{velo_field}
\end{center}
\end{figure}

On the other hand, Fig.~\ref{CO_CI_maps} shows that the two C\,{\sc i} lines do
retain a higher brightness contrast against the CMB at high redshifts than the
low-$J$ CO lines. This happens for reasons similar to those behind the
re-brightening of dust continuum distributions at high source-frame
frequencies. Indeed, the upper level energies of the two C\,{\sc i} lines
correspond to $h \nu_{10}/k_{\rm B}\sim 24$\,{\sc k} and $h \nu_{21}/k_{\rm B}
\sim 62$\,{\sc k}, which are $\gtrsim T_{\rm kin}(z)$ for cold molecular gas
(especially for the C\,{\sc i} 2--1 line). The relatively high frequencies of
the C\,{\sc i} lines make the line-CMB contrast significant as these two lines
fall on the Wien side of the SLED expected for cold H$_2$ gas (with the low-J
CO lines being decisively on its Raleigh-Jeans domain). One may wonder why the
high-J CO lines such as CO $J=6$--5, 7--6 with their $\rm h\nu_{j\rightarrow
j-1} /k_{\rm B}$ factors also $\gtrsim T_{\rm kin}(z)$ for cold molecular gas
(and thus on the Wien side of its expected CO SLED) could not be used in the
same manner as the two C\,{\sc i} lines.  This is because, unlike the low
critical densities of the two C\,{\sc i} lines ($n_{\rm crit} \sim (300-10^3)
\rm cm^{-3}$), those of high-$J$ CO lines are $\gtrsim 10^4$\,cm$^{-3}$. Thus
such CO lines will only be excited in the dense molecular gas, associated with
the typically much more compact star forming regions of galactic disks.

In Fig.~\ref{radial_gas}, we plot the radial distributions of the line emission
of CO $J$ = 1 -- 0, CO $J$ = 2 -- 1, CO $J$ = 3 -- 2, C\,{\sc i} 1 -- 0, and
C\,{\sc i} 2 -- 1 in NGC~628 at redshift of 0, 2, 4, 6 and 8, respectively.
Using the images shown in Fig.~\ref{CO_CI_maps}, we measure the average flux
density in concentric elliptical rings. We can see that the CMB effect is
dramatic, diminishing the contrast of the cold gas line emission distribution
against the CMB much more than that for the warm gas. As in the case of the
cold dust, line imaging observations would then need much longer integration
time to recover the CMB-dimmed cold molecular gas distributions in galaxies.
On the other hand they would readily recover the smaller distributions of
warmer molecular gas, thus assigning potentially much smaller sizes to the gas
disks of CMB-affected galaxies. {\it The latter would seriously underestimate
their true underlying dynamical mass, and affect the deduced M(H$_2$)/$\rm
M_{dyn}$ ratio.}

Finally, ALMA and JVLA promise not merely sensitive imaging of the dust and
H$_2$ gas mass distributions in distant gas-rich galaxies, but also the imaging
of their H$_2$ velocity fields. These, along with structural parameters such as
gas-disk scale lengths and gas and stellar mass surface densities, are
sensitive to disk instabilities in distant galaxies and yield critical
information such as $M_{\rm dyn}$.  Moreover, molecular gas velocity fields
derived from CO lines are often used in order to decide the type of a distant
heavily dust-obscured galaxy (disk versus merger).  Interferometric line
imaging observations designed to obtain high-quality H$_2$ velocity field maps
of distant galaxies are the most demanding kinds. This is because, unlike dust
continuum or velocity-integrated line brightness images, a high signal-to-noise
ratio must be achieved within narrow velocity channels in order to recover gas
velocity field information. In Fig.~\ref{velo_field} we show the effects of the
CMB on such a set of mapping observations for CO $J$=1-0, 2-1, 3-2 and C\,{\sc
i} 1-0, 2-1. As the redshift increases, the recoverable map of the molecular
gas velocity field shrinks revealing the most dramatic impact of the CMB bias
on the recoverable information from molecular line imaging observations of
distant galaxies. This in turn will strongly impact all galactic quantities
derived from such gas velocity maps such dynamical mass and the Toomre $Q$
criterion used to decide the stability of gas disks. 

Here we  must note that once  the noise associated  with any synthesis imaging
observations  and  the  cosmological  $(1+z)^{-3}$  brightness dimming factor
are taken into account, the fundamental constraints set by  the rising  CMB on
the  imaging of  cold dust  and molecular  gas distributions  of distant
galaxies  will take  effect for redshifts lower than those indicated in the
present work.  We will explore this issue using realistic  simulations of
ALMA/JVLA synthesis observations in a future paper.

\section{Summary and Conclusions}\label{summary}

We uncover and describe a fundamental constraint placed by the CMB on
cm/mm/submm imaging observations of the cold dust and molecular gas
distributions for galaxies in the distant Universe. The elevated CMB at high
redshifts dramatically diminishes the emergent continuum and line brightness
distributions of the cold gas and molecular gas. This in turn induces strong
biases on the recoverable information such as the deduced molecular gas and
dust disk distribution scale lengths, the CO-derived velocity fields, the
enclosed dynamical mass estimate, the value of the Toomre $Q$ parameter, and
the observed dust and H$_2$ gas mass distributions in galactic disks at high
redshifts.

This constraint is unique to cm, mm, and submm wavelengths as the CMB at
near-IR/optical wavelengths is negligible over the cosmic time during which
H$_2$/dust-rich galaxies are expected. Unlike the spatially-integrated effects
\cite{Papadopoulos2000ApJ,DaCunha2013ApJ,Combes1999AA} this limitation placed
by the CMB on the dust continuum and line brightness distributions cannot be
addressed simply. Nevertheless we find a unique signature that can identify
CMB-affected dust continuum or line emission brightness distributions in
high-redshift galaxies, and even recover some of the structural/dynamical
information `erased' by the elevated CMB. It consists of a nearly constant and
then rising contrast between the dust or line brightness distribution and the
CMB as the rest frame frequency of the imaging observations crosses over from
the Raleigh-Jeans to the Wien domain of the cold dust and gas S(L)ED.

The fundamental constraint set by the CMB on the imaging of cold dust and
molecular gas in the early Universe can also have a strong impact on the
cosmological census of gas-rich galaxies. This is simply because galaxies that
can be (cold-ISM)-dominated (e.g., isolated SF spirals) may be
under-represented  with respect to more (warm-ISM)-dominated galaxies
(merger/starbursts) for surveys conducted in frequencies at the Raleigh-Jeans
regime of the cold ISM emission S(L)ED.

The astronomical community was compelled to build ALMA by the natural
desire to understand the how our home, the Milky Way, has evolved
since the dawn of the cosmos.  Indeed, the design of ALMA was driven
to a significant extent by a requirement to trace the dynamics of gas
in the precursors of relatively normal galaxies like the Milky Way --
seen in the days when the Solar System was forming, in a reasonable
integration time, $\sim 1$~day. It is clear now that the dramatic
effects of the CMB will need to be considered carefully when designing
and conducting this experiment, and that studies of molecular gas and
dust in the early Universe are more complicated than we had previously
thought. The reliable detection of colossal cold, dusty structures
\cite{Ivison2012} -- lurking,
unseen, having been driven out of massive galaxies by super-winds --
was thought to be at the bleeding edge of what is currently possible,
limited by the poor sensitivity of warm single-dish telescopes in
space and interferometric studies that are `blind' on scales above
$\sim$100\,kpc. Now we understand that these technological issues may
be the least of our concerns. To faithfully discern the bulk of the
gas against the glow of the CMB, and in particular to reliably
determine its dynamics, will require a considerable investment of time
with the fully completed ALMA, with an emphasis on the gas tracers
that we have shown to be relatively immune to CMB-induced biases, such
as C\,{\sc i} lines and continuum observations at a wavelength of
about 1\,mm.

\section*{Acknowledgment}
We thank the referee for the constructive comments and suggestions, which
significantly improved the quality of this paper. Z.Y.Z thanks Huan Meng, Yu
Gao and Yong Shi for the helpful discussions.  Z.Y.Z.\ and R.J.I.\ acknowledge
support from the European Research Council in
the form of the Advanced Investigator Programme, 321302, COSMICISM.  P.P.P.\
acknowledges support from an Ernest Rutherford Fellowship.  
E.M.X.\ acknowledge financial support under the ``DeMoGas''project. The
project DeMoGas is implemented under the ``ARISTEIA'' Action of the
``Operational Programme Education and Lifelong Learning'' and is co-funded by
the European Social Fund (ESF) and National Resources.

\bibliography{cmb}
\end{document}